%% file: arxiv_manuscript.tex
\documentclass[num-refs]{wiley-article}

\usepackage{siunitx}
\usepackage{amsmath,bm,natbib,xcolor,comment,outlines,multicol,multirow}
\usepackage{graphicx}

\allowdisplaybreaks

\def\bSig\mathbf{\Sigma}

\papertype{Original Article}

\title{An IPCW adjusted Win Statistics Approach in Clinical Trials Incorporating Equivalence Margins to Define Ties}

\author[1]{Ying Cui PhD}
\author[2]{Bo Huang PhD}
\author[3]{Gaohong Dong PhD}
\author[4]{Ryuji Uozumi PhD}
\author[1]{Lu Tian ScD}

\affil[1]{Department of Biomedical Data Science, Stanford University, Stanford, California, 94304, USA}
\affil[2]{Pfizer Research \& Development, Pfizer Inc, New York, 10001, USA}
\affil[3]{Sarepta Therapeutics Inc., Cambridge, MA, 02142, USA}
\affil[4]{Department of Industrial Engineering and Economics, Tokyo Institute of Technology, Tokyo, 152-8552, Japan}

\corraddress{Ying Cui PhD, Department of Biomedical Data Science, Stanford University, Stanford, California, 94304, USA}
\corremail{yingcui@stanford.edu}

\fundinginfo{National Institutes of Health (NHLBI) [R01HL089778]}

\runningauthor{Cui et al.}

\begin{document}

\begin{frontmatter}
\maketitle

\begin{abstract}
In clinical trials, multiple outcomes of different priorities commonly occur as the patient's response may not be adequately characterized by a single outcome. Win statistics are appealing summary measures for between-group difference at more than one endpoint. When defining the result of pairwise comparisons of a time-to-event endpoint, it is desirable to allow ties to account for incomplete follow-up and not clinically meaningful difference in endpoints of interest. In this paper, we propose a class of win statistics for time-to-event endpoints with a user-specified equivalence margin. These win statistics are identifiable in the presence of right-censoring and do not depend on the censoring distribution. We then develop estimation and inference procedures for the proposed win statistics based on inverse-probability-of-censoring {weighting} (IPCW) adjustment to handle right-censoring.  We conduct extensive simulations to investigate the operational characteristics of the proposed procedure in the finite sample setting. A real oncology trial is used to illustrate the proposed approach.
\end{abstract}
\end{frontmatter}

\section{Introduction}


In clinical trials, multiple outcomes of different priorities commonly occur as the patient's response may not be adequately characterized by a single outcome.
For example, in cardiovascular disease (CVD) trials, two or more endpoints (e.g. time to CVD related death, myocardial infarction and stroke) are usually used together for quantifying the efficacy of the treatment. Conventional composite methods only analyze the time to the event that occurs first \citep{FDA2022Multiple}. Limitations of these approaches include potential loss of clinical interpretability and lack of efficiency because the ``first-occurred" event may be less important than subsequent events, which are completely ignored in the statistical analysis \citep{freemantle2003composite}. 

To overcome these limitations, a class of win statistics has recently been introduced, including the win ratio \citep{pocock2012}, the net benefit \citep{buyse2010}, and the win odds \citep{dong2019,brunner2021}.   The win statistics are defined via a contrast between the win probability for treatment and the win probability for control (denoted by $\pi_t$ and $\pi_c,$ respectively), as follows.
\begin{itemize}
    \item Win ratio: $WR = {\pi_t}/{\pi_c}$
    \item Net benefit: $NB = \pi_t - \pi_c$
    \item Win odds: $WO = \{\pi_t+0.5(1-\pi_t-\pi_c)\}/\{\pi_c+0.5(1-\pi_t-\pi_c)\}$
\end{itemize}
The win probability is defined as the chance that a randomly selected patient from a group of interest is ``better'' than a randomly selected patient from the comparison group. The specific meaning of being ``better" will be discussed later. To estimate $\pi_t$ and $\pi_c$, one may compare each patient in the treatment group with every patient in the control group, i.e.,  $N_tN_c$ pairwise comparisons, where $N_t$ and $N_c$ are the sample sizes of the treatment group and the control group, respectively \citep{luo2015,bebu2016}. Details will be provided in Sections \ref{sec:est-zeromargin} and \ref{sec:est-margin}.

There are multiple ways to define ``better" in a comparison. In this paper, we first rank all endpoints according to their clinical importance.  Then,  for each pair of patients, the comparison starts with the most important endpoint. The ``winner'' of the comparison is the patient having a better endpoint of the highest priority. If there is a tie in comparing the chosen endpoint, then the endpoint of the next highest priority is compared. This process proceeds until either a ``winner" is identified or a tie is observed for all endpoints. 
Although win statistics have the advantage in summarizing the treatment effect on multiple endpoints, it is well defined even if there is only a single endpoint. For example, if the single time-to-event endpoint in two groups follows a proportional hazards model, then the win ratio is equivalent to the reciprocal of the hazard ratio \citep{oakes2016}.

Win statistics have recently gained popularity in the designs and analyses of randomized clinical trials. 
For example, in \cite{patel2016ixmyelocel}, win statistics are utilized to assess treatment efficacy of Ixmyelocel-T for patients with ischaemic heart failure.
However, there are three major challenges to using win statistics with time-to-event type outcomes. First, the presence of right-censoring due to limited follow-up or early drop-off affects the appropriate definition of win statistics. On the one hand, according to recent ICH-E9 (R1) guidelines \citep{ICH-E9-R1}, a valid measure of treatment efficacy should not depend on the censoring distribution, which is a nuisance parameter not of our direct interest \citep{mao2022}. On the other hand, the win statistics need to be identifiable in the presence of right censoring.  Therefore, one needs to define ``better'' appropriately in comparing time to event outcomes to satisfy these two requirements. 

Second, censoring is inevitable in clinical trials with time-to-event outcomes and can result in biased inferences without appropriate correction. In the presence of a non-trivial censoring proportion, adjusting for censoring bias in estimating win statistics defined via time-to-event outcomes becomes critical \citep{dong2020a}.
\cite{dong2020b} and \cite{dong2021} introduced the inverse-probability-of-censoring weighting (IPCW) method and its extended version incorporating baseline and/or time-dependent covariates.
Those IPCW adjustment methods become complicated for scenarios with multiple time-to-event outcomes and may lead to inconsistent estimators of win probabilities, $\pi_t$ and $\pi_c$. It is interesting to observe that while there are oftentimes {non-trivial} biases in estimating $\pi_t$ and $\pi_c$ themselves, these biases occur in similar directions and magnitudes, leading to almost unbiased estimator for win statistics. Nonetheless, this phenomenon still presents a difficulty in appropriately interpreting the estimated treatment effect, in which the values of $\pi_t$ and $\pi_c$ serve as reference levels for the contrast \citep{dong2020b}.
More recently, \cite{parner2023} suggested a consistent estimator for win statistics via estimating the joint distribution of time to different endpoints of interest in the treatment and control groups separately. Since non-parametrically estimating the distribution of multivariate event times is a difficult task, this approach can only handle two endpoints and may require additional parametric or semi-parametric assumptions.

Lastly, all existing methods do not consider potential equivalence margins in defining a ``tie'', which is important since a minor difference in the selected endpoints could be caused by noise and may not be clinically important.  Specifically, an equivalence margin, denoted by $\zeta$, is a pre-specified constant that is used to determine if the event times of interest of two patients are sufficiently close to be considered a tie in the pairwise comparison.
For example, suppose that $T^{(t)}$ is the survival time of a patient in the treatment group and $T^{(c)}$ is the survival time of a patient in the control group. With the equivalence margin $\zeta$, the patient in the treatment group is better than the patient in the control group, if $T^{(t)}-T^{(c)}>\zeta.$ On the other hand, if $|T^{(c)}-T^{(t)}|\le \zeta$, we consider these two patients being tied in the comparison. 


In this paper, we propose an IPCW-adjusted method for estimating win statistics in the presence of common right-censoring while allowing for nonzero equivalence margins (i.e., zero margin is a special case of the proposed method). 
We also propose the corresponding statistical inference procedure to perform hypothesis testing and construct confidence intervals (CIs) for the chosen win statistics. 
The rest of the paper is organized as follows.
In Section 2, we propose the definition of win probabilities and the corresponding IPCW adjusted estimators, which can be used to construct estimates for win statistics. In Section 3, we describe the inference procedure based on the proposed estimators of win statistics. Sections 4 and 5 include a numerical study to investigate the finite sample performance of the proposed methods and a real data application, respectively. We conclude the paper with some remarks in Section 6.

\section{Proposed IPCW-adjusted Estimators of  Win Probabilities}


First, we introduce some necessary notation. Let ${T_l^{(t)}}$ and ${T_l^{(c)}}$ respectively denote the time to the $l$th prioritized endpoint in the treatment and control groups for $l=1,\dots,L$. Assume that there is a common censoring time for each patient, which is independent of all endpoints of interest. Denote the censoring time as $C^{(t)}$ and $C^{(c)}$ for the treatment and control groups, respectively. 
To overcome the potential identification issue due to censoring, we define win statistics by comparing truncated event times such as $T_l^{(t)}\wedge \tau$ and $T_l^{(c)}\wedge \tau,$ where $\tau$ is a chosen constant (i.e., pre-specified time horizon) such that $P(C^{(t)}\wedge C^{(c)}>\tau)>0,$ where $x\wedge y=\min(x, y)$ \citep{uno2014moving, oakes2016}. Furthermore, denote $\zeta_l$ as the pre-specified equivalence margin when comparing two event times of priority $l, l=1,\cdots, L$.   The win probabilities for the treatment and control groups are
\begin{align*}
\pi_t
&= \sum_{l=1}^L \pi_{tl}
\text{ and }
\pi_c
=\sum_{l=1}^L \pi_{cl},
\end{align*}
respectively, where 
$$
\pi_{tl} = P(T_l^{(t)}\wedge \tau > T_l^{(c)}\wedge \tau + \zeta_l, \cap_{k=0}^{l-1}{{\cal U}_k})
$$ 
and 
$$
\pi_{cl} = P(T_l^{(c)}\wedge \tau > T_l^{(t)}\wedge \tau + \zeta_l, \cap_{k=0}^{l-1}{{\cal U}_k})
$$ 
with ${\cal U}_0$ defined as the full set, and ${\cal U}_k$ as the set of tied comparisons for the $k$th endpoint, i.e.,  
$$
{\cal U}_k = \left\{ |T_k^{(t)}\wedge \tau - T_k^{(c)}\wedge \tau | \le \zeta_k \right\}.
$$

Note that $\pi_t$ and $\pi_c$ depend on the choice of  $\tau$ and equivalence margins $\{\zeta_1,\cdots, \zeta_L\}.$ Our goal is to estimate $(\pi_t, \pi_c)$ and corresponding win statistics, under the assumption of common censoring,  based on observed data consisting of 
\begin{eqnarray*}
\left\{{X_{l,i}^{(t)}} = T_{l,i}^{(t)}\wedge\tau\wedge C_i^{(t)}, {\delta_{l,i}^{(t)}} = I(T_{l,i}^{(t)}\wedge \tau \le C_i^{(t)}), i=1,\cdots, N_t, l=1, \cdots, L\right\}; \\
\left\{{X_{l,j}^{(c)}} = T_{l,j}^{(c)}\wedge\tau\wedge C_j^{(c)}, {\delta_{l,j}^{(c)}} = I(T_{l,j}^{(c)}\wedge \tau \le C_j^{(c)}), j=1, \cdots, N_c, l=1, \cdots, L\right\} 
\end{eqnarray*}
from treatment and control groups, respectively, where $I(\cdot)$ is the indicator function,
$\{({T_{1,i}^{(t)}}, T_{2,i}^{(t)}, \cdots, T_{L,i}^{(t)}, C_i^{(t)})\}_{i=1}^{N_t}$ are independent identically distributed (i.i.d.) copies of
$(T_1^{(t)}, T_2^{(t)}, \cdots, T_L^{(t)}, C^{(t)})$
and $\{({T_{1,j}^{(c)}}, T_{2,j}^{(c)}, \cdots, T_{L,j}^{(c)}, C_j^{(c)})\}_{j=1}^{N_c}$ are i.i.d. copies of
$(T_1^{(c)}, T_2^{(c)}, \cdots, T_L^{(c)}, C^{(c)}).$\\
\textbf{Remark 1 }
The selection of $\tau$ plays an important role in defining the win statistics. In general, as previously discussed in the literature for restricted mean survival time \citep{Tian2020, Huang2022}, a larger $\tau$ that is data-driven is preferred since the corresponding win statistics can capture the between-group difference in survival distribution over a wider time window.  In this paper, we propose to set $\tau$ a priori such that there is a positive proportion of patients in both groups who are still at risk at $\tau.$ In practice, one may empirically set $\tau$ as the minimum of the estimated $(1-\alpha)$ quantiles of the censoring distributions of $C^{(c)}$ and $C^{(t)},$ where $\alpha$ is a small constant such as 0.05. It can be shown that the proposed statistical inference can still perform well with this random time horizon selected based on observed data.

\subsection{Estimating win probabilities with a zero equivalence margin.} \label{sec:est-zeromargin}

We start with the simple case with $\zeta_1=\cdots=\zeta_L=0$. In this case, ${\cal U}_k$ reduces to
$$
{\cal U}_k = \left\{ T_k^{(t)}\wedge T_k^{(c)} \ge \tau \right\},
$$
i.e., a tie occurs only if two event times in a comparison are all greater than $\tau,$ the pre-specified time horizon.  The estimator for $\pi_t$ take the form:
$$
\widehat{\pi}_t = \sum_{l=1}^L \widehat{\pi}_{tl},
$$
where
\begin{align*}
\widehat{\pi}_{t1} &
=\frac{1}{N_tN_c}\sum_{i=1}^{N_t}\sum_{j=1}^{N_c}\frac{I(X_{1,i}^{(t)}>X_{1,j}^{(c)})\delta_{1,j}^{(c)}}{\widehat{G}^{(t)}(X_{1,j}^{(c)})\widehat{G}^{(c)}(X_{1,j}^{(c)})},\\
\widehat{\pi}_{tl} 
&= \frac{1}{N_tN_c}\sum_{i=1}^{N_t}\sum_{j=1}^{N_c}\frac{I(X_{l,i}^{(t)}>X_{l,j}^{(c)})\delta_{l,j}^{(c)}}{\widehat{G}^{(t)}(\tau)\widehat{G}^{(c)}(\tau)}\prod_{k=1}^{l-1} I(X_{k,i}^{(t)}=\tau,X_{k,j}^{(c)}=\tau) \text{ for }l\ge 2,
\end{align*}
where 
$\widehat{G}^{(t)}(s)$ and $\widehat{G}^{(c)}(s)$ are consistent estimators of $G^{(t)}(s)=P(C^{(t)}>s)$ and $G^{(c)}(s)=P(C^{(c)}>s)$, respectively. Specifically, $\widehat{G}^{(t)}(s)$ is the Kaplan-Meier estimator for $P(C^{(t)}>s)$ based on $\{(\widetilde{X}^{(t)}_i, \widetilde{\delta}^{(t)}_{C,i}), i=1, \cdots, N_t\}$ and $\widehat{G}^{(c)}(s)$ is the Kaplan-Meier estimator for $P(C^{(c)}>s)$ based on $\{(\widetilde{X}^{(c)}_j, \widetilde{\delta}^{(c)}_{C,j}), j=1, \cdots, N_c\},$
where $\widetilde{X}^{(t)}_i=\max\{X^{(t)}_{1,i}, \cdots, X^{(t)}_{L,i}\},$ $\widetilde{X}^{(c)}_j=\max\{X^{(c)}_{1,j}, \cdots, X^{(c)}_{L,j}\},$ $\widetilde{\delta}^{(t)}_{C,i} =I(C^{(t)}_i= \widetilde{X}^{(t)}_i)$ and $\widetilde{\delta}^{(c)}_{C,j} = I(C^{(t)}_j=\widetilde{X}^{(t)}_j).$
The estimator for $\pi_c$ takes a similar form:
$$
\widehat{\pi}_c  = \sum_{l=1}^L \widehat{\pi}_{cl} ,
$$
where
\begin{align*}
\widehat{\pi}_{c1} &
= \frac{1}{N_tN_c}\sum_{i=1}^{N_t}\sum_{j=1}^{N_c}\frac{I(X_{1,j}^{(c)}>X_{1,i}^{(t)})\delta_{1,i}^{(t)}}{\widehat{G}^{(t)}(X_{1,i}^{(t)})\widehat{G}^{(c)}(X_{1,i}^{(t)})},\\
\widehat{\pi}_{cl} 
&= \frac{1}{N_tN_c}\sum_{i=1}^{N_t}\sum_{j=1}^{N_c}\frac{I(X_{l,j}^{(c)}>X_{l,i}^{(t)})\delta_{l,i}^{(t)}}{\widehat{G}^{(t)}(\tau)\widehat{G}^{(c)}(\tau)}\prod_{k=1}^{l-1} I(X_{k,i}^{(t)}=\tau,X_{k,j}^{(c)}=\tau) \text{ for }l\ge 2.
\end{align*}
The justifications for the consistency of $\widehat{\pi}_t$ and $\widehat{\pi}_c$ can be found in Section S1 of the Supplementary Material.

We have imposed the common censoring assumption in constructing the aforementioned IPCW estimator. This assumption means that all time-to-event outcomes considered for the win statistic are subject to a common censoring mechanism. It is plausible in many clinical settings. For example, in CVD studies, it is reasonable to set the common censoring time for both the time to CVD death and the time to heart failure related hospitalization as the last follow-up date. 
However, the assumption of common censoring may not be realistic for other  disease areas, where different outcomes might be subject to different censoring processes. For example, in oncology, the censoring time for overall survival (OS) is the last follow-up date, whereas the censoring time for progression-free survival (PFS) is the last radiographical tumor assessment date. In such cases, we may consider an empirical solution to induce common censoring before analysis. Specifically, we can treat the minimal censoring times as the common censoring times for all endpoints. With this induced censoring, some observed event times may become right-censored. { This conversion is feasible only when the minimum censoring time is always known, even when the clinical event of interest occurs first. The potential loss of efficiency associated with this practice is examined in a numerical study and reported in supplementary materials.}

\noindent \textbf{Remark 2 }  The proposed IPCW adjustment is different from that in \citep{dong2021}, where the probability $\tilde{\pi}_{tl}=P\left(T_{l}^{(t)}\wedge\tau>T_{l}^{(c)}\wedge\tau,\right.$
$\left.T_{l-1}^{(t)}\wedge\tau=T_{l-1}^{(c)}\wedge\tau\right)$ is estimated by 
\begin{equation} \frac{1}{N_cN_t}\sum_{i=1}^{N_t}\sum_{j=1}^{N_c}\frac{I(X_{l,i}^{(t)}>X_{l,j}^{(c)})\delta_{l,j}^{(c)}}{\widehat{G}^{(t)}(X_{l,i}^{(t)})\widehat{G}^{(c)}(X_{l,j}^{(c)})}I(X_{l-1,i}^{(t)}=X_{l-1,j}^{(c)}=\tau), l\ge 2,
\label{eq:IPCWmistake}
\end{equation}
which is not consistent in our setting, since the expectation of the numerator of (\ref{eq:IPCWmistake}) is
\begin{align*}
& P(T_{l}^{(t)}\wedge\tau\wedge C^{(t)}>T_{l}^{(c)}\wedge \tau, T_{l-1}^{(t)}\wedge\tau\wedge C^{(t)}=T_{l-1}^{(c)}\wedge\tau\wedge C^{(c)})\\
=& P(T_{l}^{(t)}\wedge\tau>T_{l}^{(c)}\wedge \tau, T_{l-1}^{(t)}\wedge \tau=T_{l-1}^{(c)}\wedge \tau)P(C^{(t)}>\tau, C^{(c)}>\tau)\\
=& \tilde{\pi}_{tl}G^{(t)}(\tau)G^{(c)}(\tau),
\end{align*}
suggesting that the appropriate bias-correcting weight should be 
$\left\{G^{(t)}(\tau)G^{(c)}(\tau)\right\}^{-1}$ or a consistent estimator thereof.

\subsection{Estimating win probabilities with positive equivalence margins. }\label{sec:est-margin}

In the cases where positive equivalence margins are introduced to define ties, i.e., $\zeta_k> 0$, for the $k$th endpoint, we have
\begin{eqnarray*}
{\cal U}_k &=& \{ |T_k^{(t)}\wedge \tau  -T_k^{(c)}\wedge \tau |\le  \zeta_k\} \\
&=& \{ T_k^{(t)}\wedge \tau \ge T_k^{(c)}\wedge \tau - \zeta_k\}\setminus
\{ T_k^{(t)}\wedge \tau> T_k^{(c)}\wedge \tau + \zeta_k\},
\end{eqnarray*}
where $A\setminus B=A\cap B^c.$
The estimator for $\pi_t$ takes the form:
$$
\widehat{\pi}_t  = \sum_{l=1}^L \widehat{\pi}_{tl} ,
$$
where
{
\begin{align*}
\widehat{\pi}_{t1} &=
\frac{1}{N_tN_c}\sum_{i=1}^{N_t}\sum_{j=1}^{N_c}\frac{I(X_{1,i}^{(t)}>X_{1,j}^{(c)}+\zeta_1)\delta_{1,j}^{(c)}}{\widehat{G}^{(t)}(X_{1,j}^{(c)}+\zeta_1)\widehat{G}^{(c)}(X_{1,j}^{(c)})},\\
\widehat{\pi}_{tl} &= \frac{1}{N_tN_c}\sum_{i=1}^{N_t}\sum_{j=1}^{N_c}\sum_{\mathbf{s}_l\in \Omega_l} \frac{(-1)^{l+1} \prod_{k=1}^l I(X_{k,i}^{(t)}> X_{k,j}^{(c)}+s_k\zeta_k)\prod_{k=1}^l (s_k\delta_{k,j}^{(c)})}{\widehat{G}^{(t)}(\max\{(X_{k,j}^{(c)}+s_k\zeta_k), k=1, \cdots, l\})\widehat{G}^{(c)}(\max\{X_{k,j}^{(c)}, k=1,\cdots, l\})}, \text{ for } l\ge 2,
\end{align*}
where 
$\mathbf{s}_l=(s_1, \cdots, s_l)$ and $\Omega_l=\{1\}\times \{-1, 1\}^{(l-1)}.$ 
}
{ \input{SIM_revision/term_explain}}
The estimator for $\pi_c$ takes a similar form.
As for the case with zero equivalence margin, it can be shown that $\widehat{\pi}_t$ and $\widehat{\pi}_c$ are consistent estimators of win probability for treatment and control, respectively. \\ 
\textbf{Remark 3 } In finite sample settings, one may observe that $\widehat{\pi}_t+\widehat{\pi}_c>1$. In such cases, we propose to adjust the estimators for $\pi_t$ and $\pi_t$ by  
$$
\frac{\widehat{\pi}_t}{\widehat{\pi}_t+\widehat{\pi}_t+\widehat{\pi}_{tie}}~~\mbox{ and }~~\frac{\widehat{\pi}_c}{\widehat{\pi}_t+\widehat{\pi}_t+\widehat{\pi}_{tie}},
$$
respectively,  where
\begin{eqnarray*}
&&\widehat{\pi}_{tie} = \widehat{P}\{\cap_{k=1}^{L}{{\cal U}_k}\}\\
&=&\frac{1}{2N_tN_c}\sum_{i=1}^{N_t}\sum_{j=1}^{N_c}\sum_{\mathbf{s}_L\in \Omega_L^*} \frac{(-1)^{L+1} \prod_{k=1}^L I(X_{k,i}^{(t)}> X_{k,j}^{(c)}+s_k\zeta_k)\prod_{k=1}^L (s_k\delta_{k,j}^{(c)})}{\widehat{G}^{(t)}(\max\{(X_{k,j}^{(c)}+s_k\zeta_k), k=1, \cdots, L\})\widehat{G}^{(c)}(\max\{X_{k,j}^{(c)}, k=1,\cdots, L\})}\\
&+&\frac{1}{2N_tN_c}\sum_{i=1}^{N_t}\sum_{j=1}^{N_c}\sum_{\mathbf{s}_L\in \Omega_L^*} \frac{(-1)^{L+1} \prod_{k=1}^L I(X_{k,j}^{(c)}> X_{k,i}^{(t)}+s_k\zeta_k)\prod_{k=1}^L (s_k\delta_{k,i}^{(t)})}{\widehat{G}^{(t)}(\max\{(X_{k,i}^{(t)}), k=1, \cdots, L\})\widehat{G}^{(c)}(\max\{(X_{k,i}^{(t)}+s_k\zeta_k), k=1,\cdots, L\})}
\end{eqnarray*}
is an estimator for the probability of having ties, and $\Omega_l^*=\{-1, 1\}^{(L)}.$ As the sample size increases, the probability of the need for such an adjustment converges to zero.

\section{Inference Procedure based on IPCW-adjusted Win Statistics Estimators}

We can estimate win statistics based on  the estimated win probabilities for the treatment and control ($\widehat{\pi}_t$ and $\widehat{\pi}_c,$ respectively) as following:
\begin{itemize}
    \item[] Win ratio: $\widehat{WR} = {\widehat{\pi}_t}/{\widehat{\pi}_c}$.
    \item[] Win odds: $\widehat{WO} = \{\widehat{\pi}_t+0.5(1-\widehat{\pi}_t-\widehat{\pi}_c)\}/\{\widehat{\pi}_c+0.5(1-\widehat{\pi}_t-\widehat{\pi}_c)\}$.
    \item[] Net benefit: $\widehat{NB} = \widehat{\pi}_t - \widehat{\pi}_c$.
\end{itemize}
Moreover, in the Appendices B and C, we have shown that $\widehat{\pi}_t $ and $\widehat{\pi}_c $ can be written in the form of U-statistics, and
$$
\sqrt{N_t+N_c}\left( \begin{array}{c}\widehat{\pi}_t  - \pi_t \\ \widehat{\pi}_c  - \pi_c \end{array}\right)  = \frac{\sqrt{N_t+N_c}}{N_tN_c}\sum_{i=1}^{N_t}\sum_{j=1}^{N_c} \left(\begin{array}{c} K^A_{ij}\\ 
  L^A_{ij}\end{array}\right )+o_P(1),
$$
converges weakly to a bivariate Gaussian distribution $N(0, \Sigma)$ as the sample size goes to infinity.
The detailed forms of $K^A_{ij}$ and $L^A_{ij}$ are provided in Section S2 of the Supplementary Material. Furthermore, the variance covariance matrix $\Sigma$ can be consistently estimated by 
\begin{align*}
 \widehat{\Sigma}=&\frac{N_t+N_c}{N_t^2N_c(N_c-1)}\sum_{i=1}^{N_t}\sum_{j=1}^{N_c}\sum_{j'=1, j'\neq j}^{N_c}\left(\begin{array}{c}\widehat{K}_{ij}^A \\ \widehat{L}_{ij}^A \end{array}\right)\left(\begin{array}{c}\widehat{K}_{ij'}^A \\ \widehat{L}_{ij'}^A \end{array}\right)'\\
 &+ \frac{N_t+N_c}{N_t(N_t-1)N_c^2}\sum_{i=1}^{N_t}\sum_{i'=1,i'\neq i}^{N_t}\sum_{j=1}^{N_c} \left(\begin{array}{c}\widehat{K}_{ij}^A \\ \widehat{L}_{ij}^A \end{array}\right)\left(\begin{array}{c}\widehat{K}_{i'j}^A \\ \widehat{L}_{i'j}^A \end{array}\right)',
 \end{align*}
where $\widehat{K}_{ij}^A$ and $\widehat{L}_{ij}^A$ are consistent estimators of $K_{ij}^A$ and $L_{ij}^A,$ respectively. Their constructions are also provided in Section S2 of the Supplementary Material. Based on this variance estimator
$$\widehat{\Sigma}=\left(\begin{array}{cc} \widehat{\sigma}_t^2 & \widehat{\sigma}_{tc} \\ 
\widehat{\sigma}_{tc} & \widehat{\sigma}_c^2 \end{array}\right),$$ one may approximate the variance of 
$s(\widehat{\pi}_t, \widehat{\pi}_c)$ for any differentiable bivariate function $s(\cdot, \cdot)$ by
$$ \widehat{\sigma}_s^2=\frac{1}{N_c+N_t}\left\{\dot{s}_1(\widehat{\pi}_t, \widehat{\pi}_c), \dot{s}_2(\widehat{\pi}_t, \widehat{\pi}_c)\right\}\widehat{\Sigma} \left(\begin{array}{c}\dot{s}_1(\widehat{\pi}_t, \widehat{\pi}_c)\\  \dot{s}_2(\widehat{\pi}_t, \widehat{\pi}_c)\end{array}  \right),$$
where $\dot{s}_j(\cdot, \cdot)$ is the $j$th partial derivative of $s(\cdot, \cdot).$  
For $\log(WR)$, $\log(WO)$ and $NB$, the corresponding $s(\pi_{t}, \pi_{c})=\log(\pi_{t})-\log(\pi_{c})$, $\log\{1+(\pi_{t}-\pi_{c})\}-\log\{1+(\pi_{c}-\pi_{t})\}$ and $\pi_{t}-\pi_{c},$ respectively. Therefore, we may estimate the variance of $\log(\widehat{WR})$, $\log(\widehat{WO})$, and $\widehat{NB}$ by 
\begin{eqnarray*}
\widehat{\sigma}^2_{\log(WR)} &=& \frac{1}{N_c+N_t}\times\left[\frac{\widehat{\sigma}^2_t}{\widehat{\pi}_t ^2} + \frac{\widehat{\sigma}^2_c}{\widehat{\pi}_c ^2} - \frac{2\widehat{\sigma}_{tc}}{\widehat{\pi}_t \widehat{\pi}_c }\right],\\
\widehat{\sigma}^2_{\log(WO)} &=& \frac{1}{N_c+N_t}\times \frac{4(\widehat{\sigma}^2_t+\widehat{\sigma}^2_c-2\widehat{\sigma}_{tc})}{\left[1-(\widehat{\pi}_t -\widehat{\pi}_c )^2\right]^2},\\
\mbox{and}~~\widehat{\sigma}^2_{NB} &=& \frac{1}{N_c+N_t} \times (\widehat{\sigma}^2_t+\widehat{\sigma}^2_c-2\widehat{\sigma}_{tc}),
\end{eqnarray*}
respectively.  To test the treatment effect based on win statistics,  we can calculate the corresponding $z$-statistic and $p$ value.  For example, the $z$-statistic based on the win ratio is
$$
Z_{\log(WR)} = \frac{\log(\widehat{WR})}{\widehat{\sigma}_{\log(WR)}} $$
and the corresponding two-sided $p$ value can be calculated as
$p_{WR} = P\left(|N(0, 1)|>|Z_{\log(WR)}|\right)$ based on the fact that in the absence of treatment effect $Z_{\log(WR)}\sim N(0, 1)$. 
Furthermore, one may replace $\widehat{\sigma}^2_{\log(WR)}$ in the $z$-statistic by a variance estimator under the null $\pi_t=\pi_c,$ such as 
$$ \frac{4(\widehat{\sigma}^2_t+\widehat{\sigma}^2_c-2\widehat{\sigma}_{tc})}{(N_c+N_t)(1-\widehat{\pi}_{tie})^2}.$$

The test based on the win odds and the net benefit can be conducted similarly. All tests of the three win statistics are based on a contrast between $\widehat{\pi}_t $ and $\widehat{\pi}_c $ and thus are asymptotically equivalent. In other words, these three tests should have the same asymptotic power and yield very similar results, when the sample size is sufficiently {large}. 

Lastly, the $100\times(1-\alpha)\%$ CI for WR can then be constructed as
$$
\left[\widehat{WR}\times e^{ - z_{1-\alpha/2}\widehat{\sigma}_{\log(WR)}}, \widehat{WR} \times e^{ z_{1-\alpha/2}\widehat{\sigma}_{\log(WR)}} \right],
$$
where $z_{1-\alpha/2}$ denotes the $1-\alpha/2$ quantile of the standard normal distribution. Coupled with the point estimators of win probabilities and win ratio, this CI can be used to quantify the size of the difference between groups. 


\section{Simulation Studies}

In this section, we evaluate the operating characteristics of the proposed inference procedure with extensive simulations in various scenarios. Simulated data sets are generated to mimic common clinical settings. 

In the simulation study, we consider three events.  Specifically, we generate data via the following steps:
\begin{enumerate}
\item  Generate $(T_1^{(\cdot)}, T_2^{(\cdot)}, T_3^{(\cdot)})$ as 
$(F_{1(\cdot)}^{-1}\{\Phi(Z_1)\}, F_{2(\cdot)}^{-1}\{\Phi(Z_2)\}, F_{3(\cdot)}^{-1}\{\Phi(Z_3)\}),$
where $F_{l(\cdot)}(\cdot)$ denotes the selected cumulative distribution function (CDF), $F_{l(\cdot)}^{-1}(\cdot)$ is the inverse function of $F_{l(\cdot)}(\cdot)$, $\Phi(\cdot)$ denotes the CDF of $N(0,1)$, and
$$
\left(\begin{array}{c} Z_1 \\ Z_2 \\ Z_3 \end{array}\right)\sim \text{Normal}
\begin{pmatrix}
\begin{pmatrix}
0\\
0\\
0
\end{pmatrix},
\begin{pmatrix}
1 & 0.5 &  0.5\\
0.5 & 1 &  0.5\\
0.5 & 0.5 & 1
\end{pmatrix}
\end{pmatrix}.
$$
\item Generate the censoring time $C^{(\cdot)}$ independently from $F_C(\cdot)$, exponential distribution with an intensity $\lambda_C=0.02$. 
\item Calculate the observed time to the $l$th prioritized endpoint and the corresponding censoring indicator as
$$ 
{X_l^{(\cdot)}} = \min\{T_l^{(\cdot)}\wedge \tau,C^{(\cdot)}\} 
\text{ and }
{\delta_l^{(\cdot)}} = I\{T_l^{(\cdot)}\wedge \tau \le C^{(\cdot)}\}.
$$
\end{enumerate}
In the data generation, ``$\cdot$" is a generic notation, which can be ``$t$" for the treatment group and ``$c$" for the control group. In our simulation, $F_{l(t)}(\cdot)$ is the CDF of an exponential distribution with shape parameters $\lambda_l^{(t)},$ and $(\lambda_1^{(t)},\lambda_2^{(t)},\lambda_3^{(t)}) = (0.015,0.02,0.05)$.  In other words, each event time in the treatment group marginally follows an exponential distribution.

\begin{figure}
    \centering
    \includegraphics[width = \textwidth]{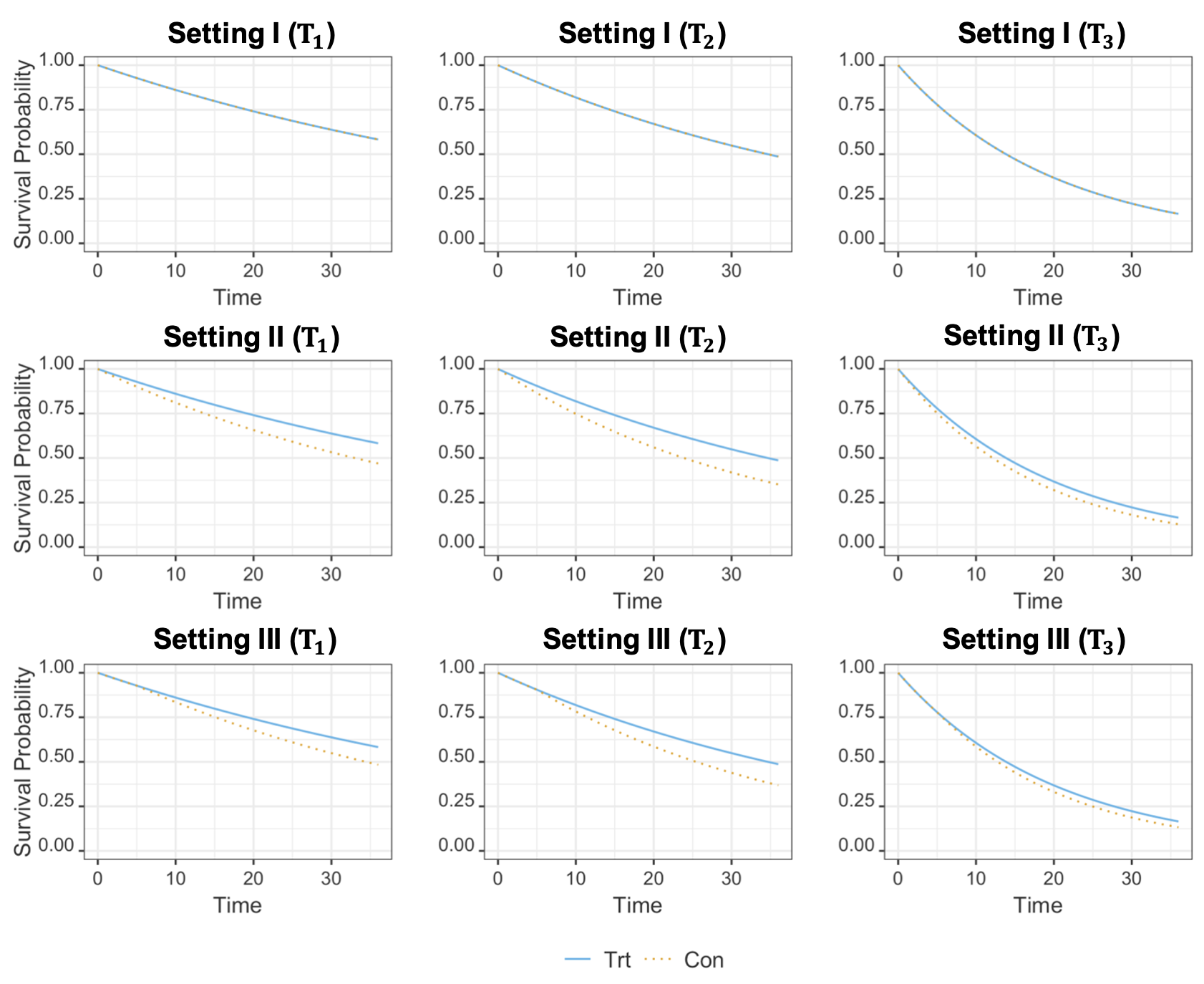}
    \caption{The survival probabilities for each endpoint in three simulation settings.}
    \label{fig:1}
\end{figure}

We consider three simulation settings. In the first setting, we mimic the null case by choosing $F_{l(c)}(\cdot)$ to be the same as $F_{l(t)}(\cdot)$ in the treatment group.
The second setting is the same as the first, except that $(\lambda_1^{(c)},\lambda_2^{(c)},\lambda_3^{(c)}) = (0.021,0.029,0.057)$, representing a proportional hazards alternative. In the third setting, $F_{l(c)}(\cdot), l=1, 2, 3$ are the CDFs of piece-wise exponential distributions:
\begin{eqnarray*}
\lambda_1^{(c)}(s) &=& 0.015 + 0.006I(s\ge 5);\\
\lambda_2^{(c)}(s) &=& 0.020 + 0.009I(s\ge 5);\\
\lambda_3^{(c)}(s) &=& 0.050 + 0.007I(s\ge 5)
\end{eqnarray*}
to investigate the performance of proposed inference procedure with a delayed treatment effect. 
Figure \ref{fig:1} presents the survival functions for each endpoint in these three settings.
Depending on the simulation setting and the endpoint of interest, the censoring rate of the truncated event times for these endpoints typically ranges from 20\% to 50\%.
We focus on estimating the win probabilities of treatment and control, and the win ratio with different combinations of $\zeta_1=\zeta_2=\zeta_3=\zeta$ and $\tau$. Their true values are calculated from a simulated data set with a sample size of 5,000,000 per group and no censoring.

Three sets of sample sizes considered include $(N_t,N_c)=(100,100)$, $(200,200)$, and $(400,400)$. 
For each simulated dataset, we estimate the win probabilities for two groups and the win ratio.  We also obtain the standard error of $\log(\widehat{WR})$. We further construct the 95\% CI of $WR$. 
Based on 5,000 simulations, we calculate the bias for each estimate, the average analytical standard error estimate for $\log(\widehat{WR}),$ the empirical standard error of $\log(\widehat{WR}),$ and the empirical coverage level of the 95\% CI for win ratio. 

The detailed results with a fixed $\zeta=0$ but different $\tau$s are summarized in Table \ref{tab:1}. 
For all settings, biases of estimating win probabilities of the treatment and control, as well as the win ratio are small relative to their true values. Also, all biases decrease as the sample size increases as expected. 
Furthermore, the empirical average of the estimated standard error and the empirical standard error are reasonably close.  
As a result, the empirical coverage levels of the 95\% CIs for WR are close to their nominal level, even with a moderate sample size of 100 per group. Moreover, as $\tau$ increases, the win probabilities in both groups increase, since there are fewer ties.

Table \ref{tab:2} summarizes the simulation results with varying $\zeta$ values. The results are very similar to those in Table \ref{tab:1}: all biases are relatively small; standard error estimates are accurate, and coverage levels of the constructed CIs are close to their nominal level. 
In addition, the win probabilities of both treatment and control decrease as $\zeta$ increases. This can be explained by the fact that a larger margin $\zeta$ induces more ties.

Furthermore, we evaluate the finite sample performance of the proposed method for hypothesis testing. To this end, a one-sided test based on the win ratio is conducted for all simulated data sets. In each setting, the significance level is set as 0.05. 
As a benchmark, we also apply the log-rank test to compare the time to the first event, whichever occurs first. The rejection rates are summarized in Table \ref{tab:3}. In the first setting without any treatment effect, the rejection rate is the same as the type I error rate, which is close to 5\% for all cases investigated. 
In the second and third simulation settings with a positive treatment effect, the rejection rate is the same as the empirical power, which increases with the sample size. Compared to the log-rank test that focuses on time to the first event, the proposed tests are substantially more powerful by aggregating information from all three events of interest. For example, in the second setting with $(\tau, \zeta) = (36, 0)$ and a sample size of 400 per group, the power is 93.1\% and 64.0\% for the WR-based test and the conventional log-rank test, respectively. 
{\input{SIM_revision/sensitivity_analyses_zeta_response}}

{ \input{SIM_revision/discussion_WR_without_IPCW}}

{ \input{SIM_revision/discussion_Weibull}}

{ \input{SIM_revision/discussion_common_censoring}}

In summary, our simulation results suggest that the proposed IPCW adjusted method perform reasonably well in finite sample cases. Statistical inferences on win statistics could be more informative than those focusing only on the time to the first event. Additionally, introducing a moderate size margin in the definition of a tie does not affect the performance of the statistical inference based on the win ratio. 


\begin{table}
  \centering
  \caption{Simulation results on bias (BIAS) in estimating the win probability for treatment and control ($\pi_t$ and $\pi_c$, respectively), and bias (BIAS) in estimating win ratio, average analytical standard error estimate of $\log(\widehat{WR})$ (ASE), empirical standard error of $\log(\widehat{WR})$ (ESE) and empirical coverage probability (CP) of 95\% CIs for win ratio across 5000 replicates with $\zeta = 0$ and $\tau\in \{18, 36\}$ in settings of exponential distributions.}
\begin{tabular}{c|c|c|c|r|c|r|c|c|c|c|c}
\hline
& & & \multicolumn{2}{c|}{$\pi_t$} & \multicolumn{2}{c|}{$\pi_c$} & \multicolumn{5}{c}{WR} \\
\cline{4-12}  Setting  & $\tau$ & $(N_t,N_c)$ & TRUE   & BIAS  & TRUE   & BIAS  & TRUE   & BIAS  & ASE   & ESE   & CP \\
\hline
I & 18 & (100,100) & 0.447 & -0.001 & 0.447 & $<$0.001 & 1.000 & 0.019 & 0.204 & 0.204 & 0.947 \\
&       & (200,200) & 0.447 & $<$0.001 & 0.447 & -0.001 & 1.000 & 0.012 & 0.144 & 0.142 & 0.946 \\
&       & (400,400) & 0.447 & $<$0.001 & 0.447 & $<$0.001 & 1.000 & 0.005 & 0.102 & 0.102 & 0.953 \\
& 36 & (100,100) & 0.493 & -0.005 & 0.493 & -0.003 & 1.000 & 0.016 & 0.204 & 0.204 & 0.949 \\
&       & (200,200) & 0.493 & -0.002 & 0.493 & -0.002 & 1.000 & 0.011 & 0.144 & 0.142 & 0.949 \\
&       & (400,400) & 0.493 & -0.001 & 0.493 & -0.001 & 1.000 & 0.005 & 0.102 & 0.099 & 0.955 \\
\hline
II & 18 & (100,100) & 0.519 & -0.001 & 0.397 & 0.001 & 1.307 & 0.025 & 0.200 & 0.200 & 0.947 \\
&       & (200,200) & 0.519 & $<$0.001 & 0.397 & -0.001 & 1.307 & 0.017 & 0.141 & 0.139 & 0.957 \\
&       & (400,400) & 0.519 & $<$0.001 & 0.397 & $<$0.001 & 1.307 & 0.005 & 0.100 & 0.099 & 0.951 \\
& 36 & (100,100) & 0.571 & -0.006 & 0.420 & -0.003 & 1.360 & 0.027 & 0.200 & 0.200 & 0.951 \\
&       & (200,200) & 0.571 & -0.003 & 0.420 & -0.003 & 1.360 & 0.018 & 0.142 & 0.138 & 0.956 \\
&       & (400,400) & 0.571 & -0.001 & 0.420 & -0.001 & 1.360 & 0.008 & 0.100 & 0.097 & 0.957 \\
\hline
III & 18 & (100,100) & 0.497 & -0.001 & 0.414 & 0.001 & 1.198 & 0.021 & 0.201 & 0.201 & 0.951 \\
&       & (200,200) & 0.497 & $<$0.001 & 0.414 & $<$0.001 & 1.198 & 0.014 & 0.142 & 0.139 & 0.957 \\
&       & (400,400) & 0.497 & -0.001 & 0.414 & $<$0.001 & 1.198 & 0.003 & 0.101 & 0.099 & 0.953 \\
& 36 & (100,100) & 0.556 & -0.006 & 0.435 & -0.003 & 1.278 & 0.024 & 0.201 & 0.200 & 0.952 \\ 
&       & (200,200) & 0.556 & -0.003 & 0.435 & -0.003 & 1.278 & 0.016 & 0.142 & 0.139 & 0.957 \\ 
&       & (400,400) & 0.556 & -0.001 & 0.435 & -0.001 & 1.278 & 0.007 & 0.101 & 0.097 & 0.956 \\
\hline
\end{tabular}
\label{tab:1}
\end{table}


\begin{table}
  \centering
  \caption{Simulation results on bias (BIAS) in estimating the win probability for treatment and control ($\pi_t$ and $\pi_c$, respectively), and bias (BIAS) in estimating win ratio, average analytical standard error estimate of $\log(\widehat{WR})$ (ASE), empirical standard error of $\log(\widehat{WR})$ (ESE) and empirical coverage probability (CP) of 95\% CIs for win ratio across 5000 replicates with different combinations of $\tau$s and $\zeta$s in different settings of exponential distributions ($N_t=N_c=200$).}
    \begin{tabular}{c|c|r|c|r|c|r|c|c|c|c|c}
    \hline
    \multirow{2}{*}{} & &  & \multicolumn{2}{c|}{$\pi_t$} & \multicolumn{2}{c|}{$\pi_c$} & \multicolumn{5}{c}{WR} \\
\cline{4-12}      Setting    & $\tau$ &   $\zeta$    & TRUE   & BIAS  & TRUE   & BIAS  & TRUE   & BIAS  & ASE   & ESE   & CP \\
    \hline
    I & 18 & 0     & 0.447 & $<$0.001 & 0.447 & -0.001 & 1.000 & 0.012 & 0.144 & 0.142 & 0.946 \\
&       & 2     & 0.424 & -0.002 & 0.424 & -0.003 & 1.000 & 0.014 & 0.150 & 0.149 & 0.944 \\
&       & 4     & 0.393 & -0.002 & 0.393 & -0.003 & 1.000 & 0.016 & 0.158 & 0.156 & 0.947 \\
&       & 6     & 0.356 & -0.001 & 0.356 & -0.003 & 1.000 & 0.019 & 0.167 & 0.166 & 0.944 \\ 
& 36 & 0     & 0.493 & -0.002 & 0.493 & -0.002 & 1.000 & 0.011 & 0.144 & 0.142 & 0.949 \\
&       & 2     & 0.487 & -0.004 & 0.487 & -0.005 & 1.000 & 0.011 & 0.145 & 0.142 & 0.952 \\
&       & 4     & 0.478 & -0.004 & 0.478 & -0.005 & 1.000 & 0.012 & 0.146 & 0.144 & 0.951 \\
&       & 6     & 0.466 & -0.004 & 0.466 & -0.004 & 1.000 & 0.012 & 0.149 & 0.147 & 0.950 \\
\hline
II & 18 & 0     & 0.519 & $<$0.001 & 0.397 & -0.001 & 1.307 & 0.017 & 0.141 & 0.139 & 0.957 \\ 
&       & 2     & 0.496 & -0.004 & 0.378 & -0.002 & 1.313 & 0.013 & 0.146 & 0.144 & 0.954 \\
&       & 4     & 0.465 & -0.003 & 0.353 & -0.003 & 1.316 & 0.017 & 0.153 & 0.151 & 0.953 \\ 
&       & 6     & 0.424 & -0.002 & 0.321 & -0.003 & 1.320 & 0.024 & 0.162 & 0.161 & 0.953 \\
& 36 & 0     & 0.571 & -0.003 & 0.420 & -0.003 & 1.360 & 0.018 & 0.142 & 0.138 & 0.956 \\ 
&       & 2     & 0.567 & -0.005 & 0.415 & -0.004 & 1.367 & 0.015 & 0.142 & 0.139 & 0.958 \\
&       & 4     & 0.559 & -0.005 & 0.408 & -0.004 & 1.371 & 0.016 & 0.143 & 0.141 & 0.956 \\
&       & 6     & 0.547 & -0.004 & 0.398 & -0.004 & 1.376 & 0.019 & 0.145 & 0.143 & 0.955 \\
\hline
III & 18 & 0     & 0.497 & $<$0.001 & 0.414 & $<$0.001 & 1.198 & 0.014 & 0.142 & 0.139 & 0.957 \\ 
&       & 2     & 0.471 & -0.003 & 0.395 & -0.002 & 1.193 & 0.011 & 0.147 & 0.145 & 0.953 \\
&       & 4     & 0.437 & -0.003 & 0.369 & -0.003 & 1.183 & 0.016 & 0.154 & 0.153 & 0.953 \\ 
&       & 6     & 0.394 & -0.001 & 0.336 & -0.003 & 1.172 & 0.023 & 0.164 & 0.163 & 0.952 \\ 
& 36 & 0     & 0.556 & -0.003 & 0.435 & -0.003 & 1.278 & 0.016 & 0.142 & 0.139 & 0.957 \\
&       & 2     & 0.550 & -0.005 & 0.430 & -0.004 & 1.280 & 0.015 & 0.143 & 0.140 & 0.959 \\ 
&       & 4     & 0.542 & -0.004 & 0.423 & -0.004 & 1.281 & 0.015 & 0.144 & 0.142 & 0.955 \\
&       & 6     & 0.530 & -0.004 & 0.413 & -0.004 & 1.284 & 0.016 & 0.146 & 0.144 & 0.954 \\ 
    \hline
    \end{tabular}
  \label{tab:2}
\end{table}


\begin{table}
  \centering 
  \caption{Simulation results for the empirical type I error rate and empirical power of the test based on win ratio estimator with IPCW adjustment (WR), naive win ratio estimator without IPCW adjustment (WR$^{(\text{no})}$) and log-rank test for comparing the time to the first occurred event (Logrank) across 5000 replicates with different combinations of $\tau$s and $\zeta$s in settings of exponential distributions.}
    \begin{tabular}{c|c|c|c|c|c|c|c|c|c|c|c}
    \hline
    & & & \multicolumn{3}{c|}{$\zeta=0$} & \multicolumn{2}{c|}{$\zeta=2$} & \multicolumn{2}{c|}{$\zeta=4$} & \multicolumn{2}{c}{$\zeta=6$} \\
\cline{4-12}   &   $\tau$    &    $(N_t,N_c)$ & WR  & WR$^{(\text{no})}$  & Logrank & WR  & WR$^{(\text{no})}$  & WR  & WR$^{(\text{no})}$  & WR & WR$^{(\text{no})}$\\
    \hline
    I & 18 & (100,100) & 0.048 & 0.051 & 0.054 & 0.048 & 0.051 & 0.052 & 0.050 & 0.051 & 0.052 \\
&       & (200,200) & 0.052 & 0.052 & 0.050 & 0.054 & 0.051 & 0.053 & 0.052 & 0.054 & 0.053 \\
&       & (400,400) & 0.046 & 0.049 & 0.054 & 0.048 & 0.048 & 0.049 & 0.049 & 0.047 & 0.047 \\
& 36 & (100,100) & 0.046 & 0.047 & 0.053 & 0.046 & 0.048 & 0.046 & 0.047 & 0.049 & 0.048 \\ 
&       & (200,200) & 0.047 & 0.051 & 0.050 & 0.047 & 0.051 & 0.048 & 0.052 & 0.052 & 0.053 \\
&       & (400,400) & 0.047 & 0.049 & 0.051 & 0.047 & 0.049 & 0.048 & 0.050 & 0.046 & 0.050 \\ 
    \hline
II & 18 & (100,100) & 0.372 & 0.381 & 0.180 & 0.359 & 0.368 & 0.344 & 0.355 & 0.333 & 0.341 \\ 
&       & (200,200) & 0.605 & 0.626 & 0.332 & 0.587 & 0.600 & 0.566 & 0.577 & 0.540 & 0.552 \\
&       & (400,400) & 0.848 & 0.856 & 0.571 & 0.834 & 0.840 & 0.817 & 0.823 & 0.786 & 0.795 \\ 
& 36 & (100,100) & 0.456 & 0.450 & 0.200 & 0.455 & 0.451 & 0.460 & 0.449 & 0.454 & 0.451 \\
&       & (200,200) & 0.716 & 0.720 & 0.376 & 0.718 & 0.724 & 0.723 & 0.725 & 0.721 & 0.719 \\
&       & (400,400) & 0.931 & 0.928 & 0.640 & 0.936 & 0.927 & 0.940 & 0.929 & 0.934 & 0.927 \\
    \hline
III & 18 & (100,100) & 0.214 & 0.186 & 0.088 & 0.197 & 0.175 & 0.184 & 0.162 & 0.169 & 0.154 \\ 
&       & (200,200) & 0.353 & 0.296 & 0.137 & 0.316 & 0.270 & 0.284 & 0.248 & 0.249 & 0.225 \\ 
&       & (400,400) & 0.556 & 0.457 & 0.214 & 0.501 & 0.415 & 0.449 & 0.376 & 0.397 & 0.348 \\ 
& 36 & (100,100) & 0.336 & 0.249 & 0.106 & 0.334 & 0.252 & 0.332 & 0.250 & 0.323 & 0.254 \\ 
&       & (200,200) & 0.538 & 0.416 & 0.187 & 0.539 & 0.415 & 0.534 & 0.413 & 0.535 & 0.418 \\ 
&       & (400,400) & 0.794 & 0.639 & 0.298 & 0.795 & 0.638 & 0.790 & 0.641 & 0.791 & 0.647 \\ 
    \hline
    \end{tabular}
  \label{tab:3}
\end{table}

\section{Example}

\begin{figure}
    \centering
    \includegraphics[width = \textwidth]{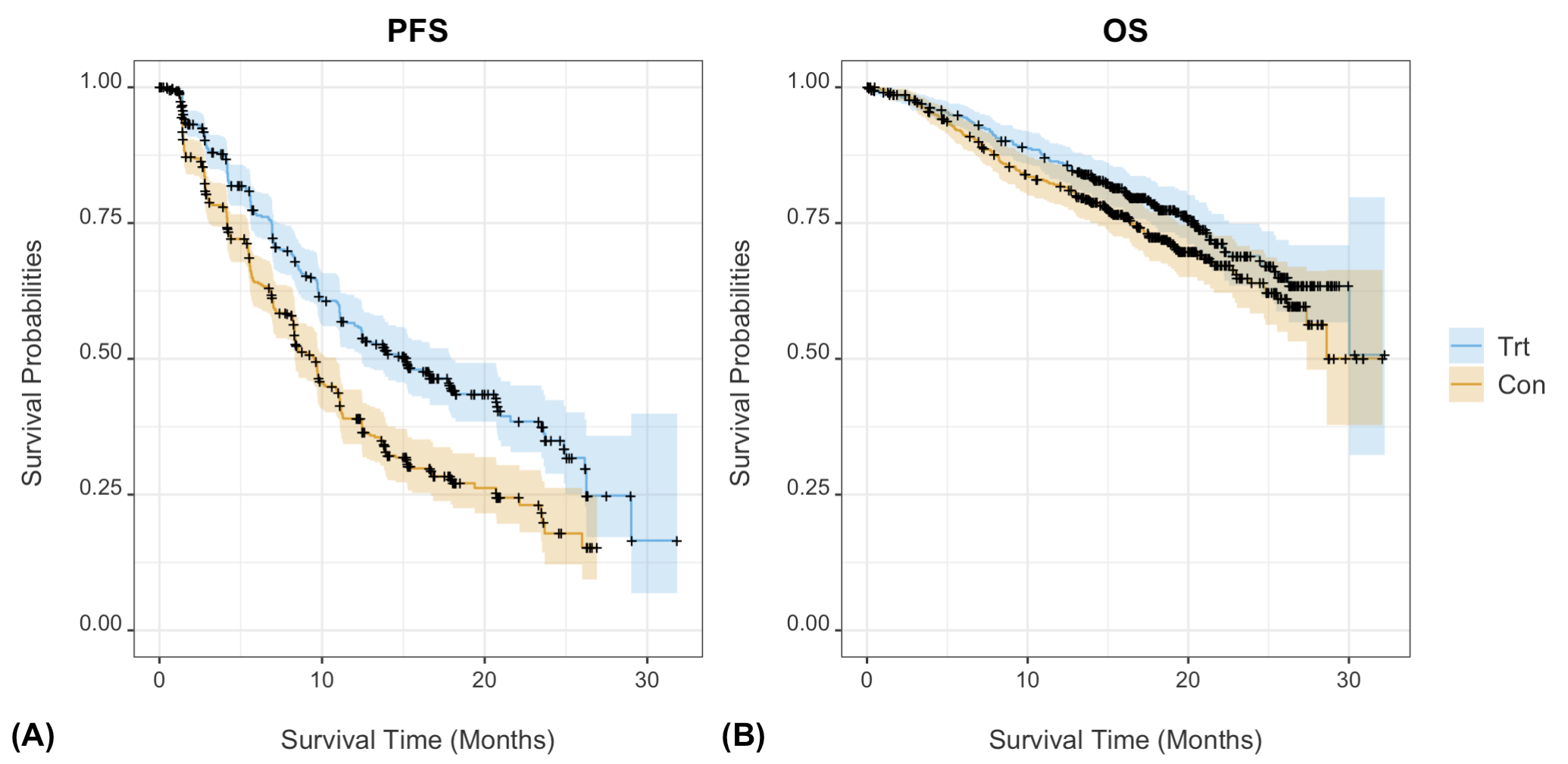}
    \caption{The Kaplan-Meier curves for PFS and OS by treatment group in JAVELIN Renal 101 trial.}
    \label{fig:2}
\end{figure}

JAVELIN Renal 101 trial \citep{motzer2019avelumab,choueiri2020} was a randomized, open-label, phase 3 study for advanced renal cell carcinoma evaluating the Avelumab + Axitinib combination therapy versus the control treatment Sunitinib.
As an illustrative example, we apply the proposed method to investigate the efficacy of Avelumab + Axitinib.  The total sample size was 886 with 442 patients randomly assigned to the treatment group and 444 patients randomized into the control group.
The endpoints of interest include OS and PFS, with the former of top priority. 
There were 109 deaths and 221 disease progressions in the treatment group, and 129 deaths and 271 disease progressions in the control group. Figure \ref{fig:2} plots the Kaplan-Meier curve for PFS and OS by treatment group, suggesting improved efficacy of the combination treatment. Also, from the graphs, we can see that the event rate of PFS is much higher than that of OS and more than 50\% of the patients experienced progression at the end of the study.

We apply the proposed method to analyze the treatment effect on OS and PFS within the time window of $[0, 24] $ months. 
The corresponding censoring rate of the truncated event time in the treatment group was about 45.7\% and 59.3\% for PFS and OS, respectively. In the control group, the censoring rate was about 37.2\% and 55.6\% for PFS and OS, respectively. For comparison purposes, we also perform conventional analyzes to examine the treatment effect on a single endpoint (either PFS or OS) based on the win ratio.

\begin{figure}
    \centering
    \includegraphics[width = \textwidth]{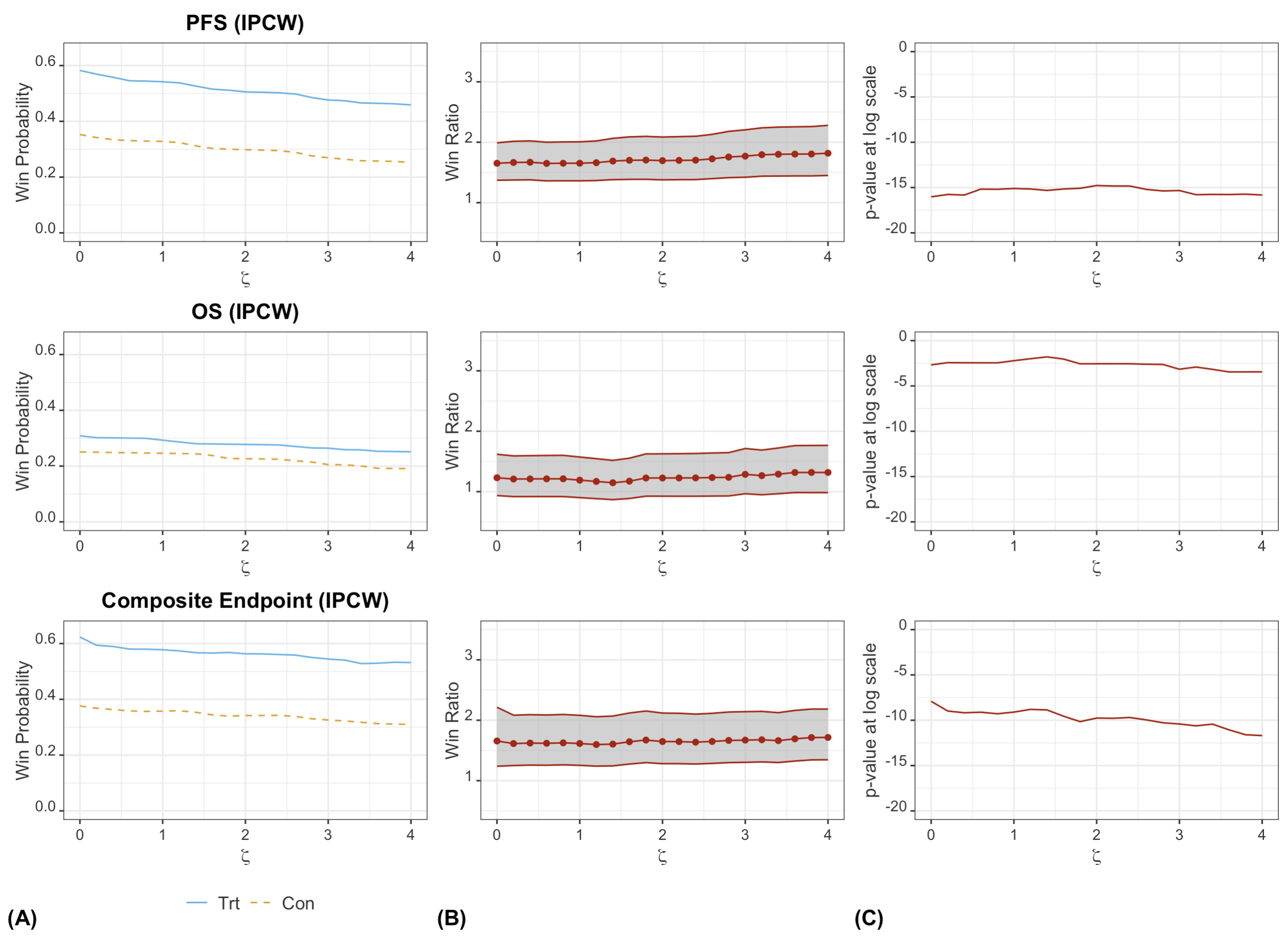}
    \caption{Results of JAVELIN Renal 101 trial. Panel A: win probabilities of treatment and control; Panel B: the estimated win ratios with 95\% CI; Panel C: $p$ value for testing the treatment effect based on the proposed IPCW-adjusted win ratio (log scale).}
    \label{fig:3}
\end{figure}

The results of these analyses are presented in Figure \ref{fig:3}. In panel A of Figure \ref{fig:3}, we present the estimated win probabilities of treatment and control with varying equivalence margin $\zeta$.  In panel B, we show the estimated win ratios with different $\zeta$ and their 95\% point-wise CI. Finally, the $p$ values based on the estimated win ratios are displayed on log scale in panel C. 
From these figures, it is evident that with the proposed method, the win probabilities for both treatment and control remain stable initially and eventually decrease as $\zeta$ increases. Meanwhile, the estimated win ratio remains almost unchanged at different values of $\zeta$. {The $p$ value tends to be robust to different values of $\zeta$.}  Specifically, when $\zeta = 0$, the estimated win ratio is 1.66 (95\% CI: [1.24,2.22]) with a $p$ value of 0.0004; when $\zeta = 2$ months, the estimated win ratio is 1.65 (95\% CI: [1.28,2.12]) with a $p$ value of $<$0.0001; and when $\zeta = 4$ months, the estimated win ratio is 1.72 (95\% CI: [1.35,2.19]) with a $p$ value of $<$0.0001. 
Moreover, we can also learn from Figure \ref{fig:3} that considering multiple endpoints can improve our power for detecting between-group differences. For example, when $\zeta = 2$ months, the estimated win ratio with respect to OS only is 1.23 (95\% CI: [0.94,1.62]) with a $p$ value of 0.069, which is less statistically significant than the aforementioned results based on both OS and PFS. 

{\input{SIM_revision/real_compare_HR_RMST}}

{\input{SIM_revision/real_WR_without_IPCW}}

\section{Discussion}

In this paper, we have proposed unbiased IPCW adjusted estimators of win statistics for time-to-event outcomes of different priorities in the presence of right-censoring, allowing non-zero equivalence margins in defining ties in pairwise comparisons. We have also proposed valid statistical inference procedures for hypothesis testing and confidence interval constructions. This paper has substantially extended the previous work by \cite{dong2020b,dong2021} and \cite{parner2023}. Based on extensive simulations and a real case study, it appears that the proposed estimators perform well in finite sample settings, and the analysis based on multiple endpoints tends to be more informative than those relying on only a single endpoint.  We encourage clinical trialists to use the proposed unbiased estimators for win statistics in future practice. 
{ The R package \texttt{WINS}, which provides implementation of functions related to the proposed method, is publicly available on CRAN at \url{https://cran.r-project.org/web/packages/WINS/index.html}.}

When the censoring time differs for different time-to-event outcomes, we propose to first induce a common censoring time, which can be used in the subsequent statistical analysis predicated on the IPCW adjustment. In special cases, one may be able to estimate the joint distribution of endpoint-specific censoring times and apply the IPCW adjustment directly to avoid the efficiency loss associated with induced common censoring. 

As shown in our simulations, the win probabilities decrease as $\zeta$ increases, because a larger margin $\zeta$ induces more ties. {\input{SIM_revision/discussionnew1}} 
In general, the margin $\zeta$ is specific to the disease indication and the study endpoints. For the same endpoint (e.g., time to death), one disease area may require a smaller margin than other indications. Therefore, it is important to align the margin among all study stakeholders to ensure that $\zeta$ is appropriate clinically and statistically. {\input{SIM_revision/discussion_choosing_zeta}}

Lastly, additional considerations are needed in using win statistics with multiple endpoints when the endpoints of interest are of different types. For example, one may be interested in survival time and tumor response, which is a binary response status. In addition, the comparison between two patients does not have to be conducted by sequentially comparing each endpoint. One can design a set of more complex rules to determine the comparison result simultaneously considering all endpoints as in \cite{claggett2015treatment}.   In such cases, the estimation and corresponding inference methods need to be tailored accordingly. There is no universal inference procedure to handle any rules that define wins within the framework of win statistics.

\section*{Acknowledgements}

This work is partially supported by the National Institutes of Health (NHLBI) [R01HL089778].


\section*{Conflicts of Interest}

The authors declare no conflicts of interest.

\section*{Data Availability Statement}

The data used are proprietary to Pfizer Inc. and was used for illustration purpose only.


\printendnotes

\bibliography{WINS}

\end{document}


\maketitle

\section{Consistency of the estimate for the win probabilities when $\zeta=0$}

Denote $\bm{T}=(\{T_l^{(t)}\}_{l=1}^L,\{T_l^{(c)}\}_{l=1}^L)$, and $\bm{C}=(C^{(t)},C^{(c)})$.
The justification for unbiasedness is as follows. Specifically, when $l=1$, we have
\begin{eqnarray*}
&&E\left\{\frac{I(X_1^{(t)}>X_1^{(c)})\delta_1^{(c)}}{G^{(t)}(X_1^{(c)})G^{(c)}(X_1^{(c)})}\right\}
= E_{\bm{T}}\left\{E_{\bm{C}}\left[\frac{I(X_1^{(t)}>X_1^{(c)})\delta_1^{(c)}}{G^{(t)}(X_1^{(c)})G^{(c)}(X_1^{(c)})}\mid \bm{T}\right]\right\}\\
&=& E_{\bm{T}}\left\{E_{\bm{C}}\left[\frac{I(T_1^{(t)}\wedge \tau>T_1^{(c)}\wedge \tau)I(C^{(t)}>T_1^{(c)}\wedge \tau)I(C^{(c)}\ge T_1^{(c)}\wedge \tau)}{G^{(t)}(T_1^{(c)}\wedge \tau)G^{(c)}(T_1^{(c)}\wedge \tau)}\mid \bm{T}\right]\right\}\\
&=& E_{\bm{T}}\left\{I(T_1^{(t)}\wedge \tau>T_1^{(c)}\wedge \tau)\cdot\frac{G^{(t)}(T_1^{(c)}\wedge \tau)G^{(c)}(T_1^{(c)}\wedge \tau)}{G^{(t)}(T_1^{(c)}\wedge \tau)G^{(c)}(T_1^{(c)}\wedge \tau)}\right\}\\
&=& P(T_1^{(t)}\wedge \tau>T_1^{(c)}\wedge \tau).
\end{eqnarray*}
Similarly, we can show that
\begin{eqnarray*}
&&E\left\{\frac{I(X_1^{(c)}>X_1^{(t)})\delta_1^{(t)}}{G^{(t)}(X_1^{(t)})G^{(c)}(X_1^{(t)})}\right\}
= P(T_1^{(c)}\wedge \tau>T_1^{(t)}\wedge \tau).
\end{eqnarray*}
For $l\ge 2$, we have
\begin{eqnarray*}
&&E\left\{\frac{I(X_l^{(t)}>X_l^{(c)})\delta_l^{(c)}}{G^{(t)}(\tau)G^{(c)}(\tau)}\prod_{k=1}^{l-1} I(X_k^{(t)}=\tau,X_k^{(c)}=\tau)\right\}\\
&=& E_{\bm{T}}\left\{E_{\bm{C}}\left[\frac{I(X_l^{(t)}>X_l^{(c)})\delta_l^{(c)}}{G^{(t)}(\tau)G^{(c)}(\tau)}\prod_{k=1}^{l-1} I(X_k^{(t)}=\tau,X_k^{(c)}=\tau)\mid \bm{T}\right]\right\}\\
&=& E_{\bm{T}}\left\{E_{\bm{C}}\left[\frac{I(T_l^{(t)}\wedge \tau>T_l^{(c)}\wedge \tau)I(C^{(t)}>T_l^{(c)}\wedge \tau)I(C^{(c)}\ge T_l^{(c)}\wedge \tau)}{G^{(t)}(\tau)G^{(c)}(\tau)}\right.\right.\\
&&\left.\left.\times\prod_{k=1}^{l-1} I(T_k^{(t)}\wedge \tau=\tau,T_k^{(c)}\wedge \tau=\tau,C^{(t)}\ge\tau,C^{(c)}\ge\tau)\mid \bm{T}\right]\right\}\\
&=& E_{\bm{T}}\left\{I(T_l^{(t)}\wedge \tau>T_l^{(c)}\wedge \tau)\times\prod_{k=1}^{l-1} I(T_k^{(t)}\wedge \tau=\tau,T_k^{(c)}\wedge \tau=\tau)\right.\\
&&\left.\times E_{\bm{C}}\left[\frac{I(C^{(t)}\ge\tau)I(C^{(c)}\ge\tau)}{G^{(t)}(\tau)G^{(c)}(\tau)}
\mid \bm{T}\right]\right\}\\
&=& E_{\bm{T}}\left\{I(T_l^{(t)}\wedge \tau>T_l^{(c)}\wedge \tau)\times\prod_{k=1}^{l-1} I(T_k^{(t)}\wedge \tau=\tau,T_k^{(c)}\wedge \tau=\tau)\times \left[\frac{G^{(t)}(\tau)G^{(c)}(\tau)}{G^{(t)}(\tau)G^{(c)}(\tau)}\right]\right\}\\
&=& P(T_l^{(c)}\wedge \tau > T_l^{(t)}\wedge \tau, \cap_{k=1}^{l-1}{{\cal U}_k}).
\end{eqnarray*}
Similarly, we can show that
\begin{eqnarray*}
&&E\left\{\frac{I(X_l^{(c)}>X_l^{(t)})\delta_l^{(t)}}{G^{(t)}(\tau)G^{(c)}(\tau)}\prod_{k=1}^{l-1} I(X_k^{(t)}=\tau,X_k^{(c)}=\tau)\right\} = P(T_l^{(t)}\wedge \tau > T_l^{(c)}\wedge \tau, \cap_{k=1}^{l-1}{{\cal U}_k}).
\end{eqnarray*}
Combining all the above results, we can easily show that $\widehat{\pi}_t(\tau)$ and $\widehat{\pi}_c(\tau)$ are unbiased estimators of $\pi_t$ and $\pi_c$, respectively.

\section{U-statistics representation of win probability estimators}
\label{sec:method3}
In this section, we discuss how to calculate $K^A_{ij}$ and $L^A_{ij}$ and their estimators. 
Note that the proposed estimator, $\widehat{\pi}_t=\sum_{l=1}^L \widehat{\pi}_{tl},$ can be expressed as the summation of $2^L-1$ terms taking a general form of
$$
\widehat{P}_0 \doteq \frac{1}{N_tN_c}\sum_{i=1}^{N_t}\sum_{j=1}^{N_c}\widehat{P}_{ij}=\frac{1}{N_tN_c}\sum_{i=1}^{N_t}\sum_{j=1}^{N_c}\frac{f(\bm{X}_{i}^{(t)},\bm{X}_{j}^{(c)},\bm{\delta}_{i}^{(t)},\bm{\delta}_{j}^{(c)})}{\widehat{G}^{(t)}[g_1(\bm{X}_{j}^{(c)})]\widehat{G}^{(c)}[g_{2}(\bm{X}_{j}^{(c)})]},
$$
where $\bm{X}_{i}^{(t)}\doteq (X_{1,i}^{(t)},\dots,X_{L,i}^{(t)})$, $\bm{X}_{j}^{(c)}\doteq (X_{1,j}^{(c)},\dots,X_{L,j}^{(c)})$, 
$\bm{\delta}_{i}^{(t)}\doteq (\delta_{1,i}^{(t)},\dots,\delta_{L,i}^{(t)})$ and $\bm{\delta}_{j}^{(c)}\doteq (\delta_{1,j}^{(c)},\dots,\delta_{L,j}^{(c)})$,
and $f(\cdot)$, $g_{1}(\cdot)$ and $g_{2}(\cdot)$ denote some general functions. It can be shown that this type of statistics asymptotically takes the form of a two-sample U-statistic. Specifically, let
\begin{eqnarray*}
P_0 = E(P_{ij})=E\left\{\frac{f(\bm{X}_{i}^{(t)},\bm{X}_{j}^{(c)},\bm{\delta}_{i}^{(t)},\bm{\delta}_{j}^{(c)})}{G^{(t)}[g_{1}(\bm{X}_{j}^{(c)})]G^{(c)}[g_{2}(\bm{X}_{j}^{(c)})]}\right\}.
\end{eqnarray*}
In Section S3, we have shown that 
\begin{eqnarray*}
&&\frac{\sqrt{N_t+N_c}}{N_tN_c}\sum_{i=1}^{N_t}\sum_{j=1}^{N_c}\left\{\frac{f(\bm{X}_{i}^{(t)},\bm{X}_{j}^{(c)},\bm{\delta}_{i}^{(t)},\bm{\delta}_{j}^{(c)})}{\widehat{G}^{(t)}[g_{1}(\bm{X}_{j}^{(c)})]\widehat{G}^{(c)}[g_{2}(\bm{X}_{j}^{(c)})]} - P_0\right\}
= \frac{\sqrt{N_t+N_c}}{N_tN_c}\sum_{i=1}^{N_t}\sum_{j=1}^{N_c}\xi_{ij}+o_p(1),
\end{eqnarray*}
where 
\begin{eqnarray*}
\xi_{ij} &\doteq& (P_{ij} - P_0) + \int_0^{\infty}\frac{E\left\{f(\bm{X}_{k}^{(t)},\bm{X}_{j}^{(c)},\bm{\delta}_{k}^{(t)},\bm{\delta}_{j}^{(c)})\mid \bm{X}_j^{(c)}, \bm{\delta}_j^{(c)}\right\}I[g_{1}(\bm{X}_{j}^{(c)})>s]}{G^{(t)}[g_{1}(\bm{X}_{j}^{(c)})]G^{(c)}[g_{2}(\bm{X}_{j}^{(c)})]}\cdot\frac{dM_i^{G^{(t)}}(s)}{y^{(t)}(s)}\\
&+& \int_0^{\infty}E\left\{\frac{f(\bm{X}_{i}^{(t)},\bm{X}_{m}^{(c)},\bm{\delta}_{i}^{(t)},\bm{\delta}_{m}^{(c)})I[g_{2}(\bm{X}_{m}^{(c)})>s]}{G^{(t)}[g_{1}(\bm{X}_{m}^{(c)})]G^{(c)}[g_{2}(\bm{X}_{m}^{(c)})]} \biggm| \bm{X}_i^{(t)}, \bm{\delta}_i^{(t)}\right\}\cdot\frac{dM_j^{G^{(c)}}(s)}{y^{(c)}(s)},
\end{eqnarray*}
$y^{(t)}(s) = P(\widetilde{X}^{(t)}\ge s)$, $y^{(c)}(s) = P(\widetilde{X}^{(c)}\ge s)$, and 
\begin{align*}
M_{C_i^{(t)}}(s)&=I(\widetilde{X}_i^{(t)}\le s)\widetilde{\delta}_{C,i}^{(t)}-\int_0^s I(\widetilde{X}_i^{(t)}\ge u)\lambda_{C^{(t)}}(u)du\\
M_{C_j^{(c)}}(s)&= I(\widetilde{X}_j^{(c)}\le s)\widetilde{\delta}_{C,j}^{(c)}-\int_0^s I(\widetilde{X}_j^{(c)}\ge u)\lambda_{C^{(c)}}(u)du
\end{align*}
denote the corresponding martingale process of censoring time in the treatment and control groups, respectively. Here $\lambda_{C^{(t)}}(\cdot)$ and $\lambda_{C^{(c)}}(\cdot)$ are hazard functions for censoring time $C^{(t)}$ and $C^{(c)},$ respectively. Therefore, $K_{ij}^{A}$ is simply the summation of $\xi_{ij}$s corresponding to the terms in $\widehat{\pi}_{t}$s. Furthermore, $\xi_{ij}$ can be consistently estimated by 
\begin{eqnarray*}
\widehat{\xi}_{ij} &\doteq& (\widehat{P}_{ij} - \widehat{P}_{0}) + \int_0^{\infty}\left\{\frac{1}{N_t}\sum_{k=1}^{N_t}\frac{f(\bm{X}_{k}^{(t)},\bm{X}_{j}^{(c)},\bm{\delta}_{k}^{(t)},\bm{\delta}_{j}^{(c)})I[g_{1}(\bm{X}_{j}^{(c)})>s]}{\widehat{G}^{(t)}[g_{1}(\bm{X}_{j}^{(c)})]\widehat{G}^{(c)}[g_{2}(\bm{X}_{j}^{(c)})]}\right\}\cdot\frac{d\widehat{M}_{C_i^{(t)}}(s)}{\widehat{y}^{(t)}(s)}\\
&+& \int_0^{\infty}\left\{\frac{1}{N_c}\sum_{m=1}^{N_c}\frac{f(\bm{X}_{i}^{(t)},\bm{X}_{m}^{(c)},\bm{\delta}_{i}^{(t)},\bm{\delta}_{m}^{(c)})I[g_{2}(\bm{X}_{m}^{(c)})>s]}{\widehat{G}^{(t)}[g_{1}(\bm{X}_{m}^{(c)})]\widehat{G}^{(c)}[g_{2}(\bm{X}_{m}^{(c)})]}\right\}\cdot\frac{d\widehat{M}_{C_j^{(c)}}(s)}{\widehat{y}^{(c)}(s)},
\end{eqnarray*}
where 
\begin{align*}
\widehat{M}_{C_i^{(t)}}(s)&=I(\widetilde{X}_i^{(t)}\le s)\widetilde{\delta}_{C,i}^{(t)}-\int_0^s I(\widetilde{X}_i^{(t)}\ge u)d\left[-\log\{\widehat{G}^{(t)}(u)\}\right],\\
\widehat{M}_{C_j^{(c)}}(s)&= I(\widetilde{X}_j^{(c)}\le s)\widetilde{\delta}_{C,j}^{(c)}-\int_0^s I(\widetilde{X}_j^{(c)}\ge u)d\left[-\log\{\widehat{G}^{(c)}(u)\}\right],\\
\widehat{y}^{(t)}(s)&=\frac{1}{N_t}\sum_{i=1}^{N_t} I(\widetilde{X}^{(t)}_i\ge s)~~\mbox{and}~~\widehat{y}^{(c)}(s)=\frac{1}{N_c}\sum_{j=1}^{N_c} I(\widetilde{X}^{(c)}_j\ge s).
\end{align*}

Finally, $\widehat{K}_{ij}^{A}$ can be constructed as the summation of $\widehat{\xi}_{ij}$s.
We may express $L_{ij}^{A}$ as the summation of $\eta_{ij}$s, whose forms are given in this Supplementary Material. Similarly, we may construct $\widehat{\eta}_{ij}$s and then $\widehat{L}_{ij}^{A}$ accordingly. More details are provided in the Section S3.

\section{First order representation of the estimate for win probabilities}


From \cite{pepe1991}, we have
\begin{eqnarray*}
N_t^{1/2}\left[\widehat{G}^{(t)}(u) - G^{(t)}(u)\right] &=& - N_t^{-1/2}\sum_{i=1}^n G^{(t)}(u) \int_0^{u} y^{(t)}(s)^{-1}dM_i^{G^{(t)}}(s)+o_p(1),\\
N_c^{1/2}\left[\widehat{G}^{(c)}(u) - G^{(c)}(u)\right] &=& - N_c^{-1/2}\sum_{j=1}^n G^{(c)}(u)\int_0^{u} y^{(c)}(s)^{-1}dM_j^{G^{(c)}}(s)+o_p(1),
\end{eqnarray*}
where $o_p(1)$ terms are uniform over $u\in [0, \tau].$
Thus for any $\xi_{ij}$, we have
{\small
\begin{eqnarray*}
&& \frac{\sqrt{N_t+N_c}}{N_tN_c}\sum_{i=1}^{N_t}\sum_{j=1}^{N_c} \left\{\frac{f(\bm{X}_{i}^{(t)},\bm{X}_{j}^{(c)},\bm{\delta}_{i}^{(t)},\bm{\delta}_{j}^{(c)})}{\widehat{G}^{(t)}[g_{1}(\bm{X}_{j}^{(c)})]\widehat{G}^{(c)}[g_{2}(\bm{X}_{j}^{(c)})]} - P_0\right\}\\
&=& \frac{\sqrt{N_t+N_c}}{N_tN_c}\sum_{i=1}^{N_t}\sum_{j=1}^{N_c} (P_{ij} - P_0) \\
&-& \frac{\sqrt{N_t+N_c}}{N_tN_c}\sum_{i=1}^{N_t}\sum_{j=1}^{N_c} \frac{f(\bm{X}_{i}^{(t)},\bm{X}_{j}^{(c)},\bm{\delta}_{i}^{(t)},\bm{\delta}_{j}^{(c)})}{\{G^{(t)}[g_{1}(\bm{X}_{j}^{(c)})]\}^2G^{(c)}[g_{2}(\bm{X}_{j}^{(c)})]}\left\{\widehat{G}^{(t)}[g_{1}(\bm{X}_{j}^{(c)})]-G^{(t)}[g_{1}(\bm{X}_{j}^{(c)})]\right\}\\
&-& \frac{\sqrt{N_t+N_c}}{N_tN_c}\sum_{i=1}^{N_t}\sum_{j=1}^{N_c} \frac{f(\bm{X}_{i}^{(t)},\bm{X}_{j}^{(c)},\bm{\delta}_{i}^{(t)},\bm{\delta}_{j}^{(c)})}{G^{(t)}[g_{1}(\bm{X}_{j}^{(c)})]\{G^{(c)}[g_{2}(\bm{X}_{j}^{(c)})]\}^2}\left\{\widehat{G}^{(c)}[g_{2}(\bm{X}_{j}^{(c)})]-G^{(c)}[g_{2}(\bm{X}_{j}^{(c)})]\right\}+o_p(1)\\
&=& \frac{\sqrt{N_t+N_c}}{N_tN_c}\sum_{i=1}^{N_t}\sum_{j=1}^{N_c}(P_{ij} - P_0) \\
&+& \frac{\sqrt{N_t+N_c}}{N_tN_c}\sum_{i=1}^{N_t}\sum_{j=1}^{N_c}\frac{1}{\sqrt{N_t}}\cdot\frac{f(\bm{X}_{i}^{(t)},\bm{X}_{j}^{(c)},\bm{\delta}_{i}^{(t)},\bm{\delta}_{j}^{(c)})}{G^{(t)}[g_{1}(\bm{X}_{j}^{(c)})]G^{(c)}[g_{2}(\bm{X}_{j}^{(c)})]}\cdot\frac{1}{\sqrt{N_t}}\sum_{k=1}^{N_t}\int_0^{\infty}\frac{I[g_{1}(\bm{X}_{j}^{(c)})>s]dM_k^{G^{(t)}}(s)}{y^{(t)}(s)}\\
&+& \frac{\sqrt{N_t+N_c}}{N_tN_c}\sum_{i=1}^{N_t}\sum_{j=1}^{N_c}\frac{1}{\sqrt{N_c}}\cdot\frac{f(\bm{X}_{i}^{(t)},\bm{X}_{j}^{(c)},\bm{\delta}_{i}^{(t)},\bm{\delta}_{j}^{(c)})}{G^{(t)}[g_{1}(\bm{X}_{j}^{(c)})]G^{(c)}[g_{2}(\bm{X}_{j}^{(c)})]}\cdot\frac{1}{\sqrt{N_c}}\sum_{m=1}^{N_c}\int_0^{\infty}\frac{I[g_{2}(\bm{X}_{j}^{(c)})>s]dM_m^{G^{(c)}}(s)}{y^{(c)}(s)}+o_p(1)\\
&=& \frac{\sqrt{N_t+N_c}}{N_tN_c}\sum_{i=1}^{N_t}\sum_{j=1}^{N_c}(P_{ij} - P_0) \\
&+& \frac{\sqrt{N_t+N_c}}{N_tN_c}\sum_{i=1}^{N_t}\sum_{j=1}^{N_c}
\int_0^{\infty}\left\{\frac{1}{N_t}\sum_{k=1}^{N_t}\frac{f(\bm{X}_{k}^{(t)},\bm{X}_{j}^{(c)},\bm{\delta}_{k}^{(t)},\bm{\delta}_{j}^{(c)})I[g_{1}(\bm{X}_{j}^{(c)})>s]}{G^{(t)}[g_{1}(\bm{X}_{j}^{(c)})]G^{(c)}[g_{2}(\bm{X}_{j}^{(c)})]}\right\}\cdot\frac{dM_i^{G^{(t)}}(s)}{y^{(t)}(s)}\\
&+& \frac{\sqrt{N_t+N_c}}{N_tN_c}\sum_{i=1}^{N_t}\sum_{j=1}^{N_c}
\int_0^{\infty}\left\{\frac{1}{N_c}\sum_{m=1}^{N_c}\frac{f(\bm{X}_{i}^{(t)},\bm{X}_{m}^{(c)},\bm{\delta}_{i}^{(t)},\bm{\delta}_{m}^{(c)})I[g_{2}(\bm{X}_{m}^{(c)})>s]}{G^{(t)}[g_{1}(\bm{X}_{m}^{(c)})]G^{(c)}[g_{2}(\bm{X}_{m}^{(c)})]}\right\}\cdot\frac{dM_j^{G^{(c)}}(s)}{y^{(c)}(s)}+o_p(1)\\
&=& \frac{\sqrt{N_t+N_c}}{N_tN_c}\sum_{i=1}^{N_t}\sum_{j=1}^{N_c}(P_{ij} - P_0) \\
&+& \frac{\sqrt{N_t+N_c}}{N_tN_c}\sum_{i=1}^{N_t}\sum_{j=1}^{N_c}
\int_0^{\infty}\frac{E\left\{f(\bm{X}_{k}^{(t)},\bm{X}_{j}^{(c)},\bm{\delta}_{k}^{(t)},\bm{\delta}_{j}^{(c)})\mid \bm{X}_j^{(c)}, \bm{\delta}_j^{(c)}\right\}I[g_{1}(\bm{X}_{j}^{(c)})>s]}{G^{(t)}[g_{1}(\bm{X}_{j}^{(c)})]G^{(c)}[g_{2}(\bm{X}_{j}^{(c)})]}\cdot\frac{dM_i^{G^{(t)}}(s)}{y^{(t)}(s)}\\
&+& \frac{\sqrt{N_t+N_c}}{N_tN_c}\sum_{i=1}^{N_t}\sum_{j=1}^{N_c}
\int_0^{\infty}E\left\{\frac{f(\bm{X}_{i}^{(t)},\bm{X}_{m}^{(c)},\bm{\delta}_{i}^{(t)},\bm{\delta}_{m}^{(c)})I[g_{2}(\bm{X}_{m}^{(c)})>s]}{G^{(t)}[g_{1}(\bm{X}_{m}^{(c)})]G^{(c)}[g_{2}(\bm{X}_{m}^{(c)})]} \biggm| \bm{X}_i^{(t)}, \bm{\delta}_i^{(t)}\right\}\cdot\frac{dM_j^{G^{(c)}}(s)}{y^{(c)}(s)}+o_p(1).
\end{eqnarray*}
}
Therefore, we have
 $$\frac{\sqrt{N_t+N_c}}{N_tN_c}\sum_{i=1}^{N_t}\sum_{j=1}^{N_c}\left\{\frac{f(\bm{X}_{i}^{(t)},\bm{X}_{j}^{(c)},\bm{\delta}_{i}^{(t)},\bm{\delta}_{j}^{(c)})}{\widehat{G}^{(t)}[g_{1}(\bm{X}_{j}^{(c)})]\widehat{G}^{(c)}[g_{2}(\bm{X}_{j}^{(c)})]} - P_0\right\}=\frac{\sqrt{N_t+N_c}}{N_tN_c}\sum_{i=1}^{N_t}\sum_{j=1}^{N_c} \xi_{ij}+o_p(1),$$
where
\begin{eqnarray*}
\xi_{ij} &\doteq& (P_{ij} - P_0) + \int_0^{\infty}\frac{E\left\{f(\bm{X}_{k}^{(t)},\bm{X}_{j}^{(c)},\bm{\delta}_{k}^{(t)},\bm{\delta}_{j}^{(c)})\mid \bm{X}_j^{(c)}, \bm{\delta}_j^{(c)}\right\}I[g_{1}(\bm{X}_{j}^{(c)})>s]}{G^{(t)}[g_{1}(\bm{X}_{j}^{(c)})]G^{(c)}[g_{2}(\bm{X}_{j}^{(c)})]}\cdot\frac{dM_i^{G^{(t)}}(s)}{y^{(t)}(s)}\\
&+& \int_0^{\infty}E\left\{\frac{f(\bm{X}_{i}^{(t)},\bm{X}_{m}^{(c)},\bm{\delta}_{i}^{(t)},\bm{\delta}_{m}^{(c)})I[g_{2}(\bm{X}_{m}^{(c)})>s]}{G^{(t)}[g_{1}(\bm{X}_{m}^{(c)})]G^{(c)}[g_{2}(\bm{X}_{m}^{(c)})]} \biggm| \bm{X}_i^{(t)}, \bm{\delta}_i^{(t)}\right\}\cdot\frac{dM_j^{G^{(c)}}(s)}{y^{(c)}(s)}.
\end{eqnarray*}
Similarly, we can write 
\begin{eqnarray*}
\eta_{ij} &\doteq& (P_{ij} - P_0) + \int_0^{\infty}E\left\{\frac{f(\bm{X}_{k}^{(t)},\bm{X}_{j}^{(c)},\bm{\delta}_{k}^{(t)},\bm{\delta}_{j}^{(c)})I[g_{1}(\bm{X}_{k}^{(t)})>s]}{G^{(t)}[g_{1}(\bm{X}_{k}^{(t)})]G^{(c)}[g_{2}(\bm{X}_{k}^{(t)})]}
\biggm| \bm{X}_{j}^{(t)},\bm{\delta}_{j}^{(t)}\right\}
\cdot\frac{dM_i^{G^{(t)}}(s)}{y^{(t)}(s)}\\
&+& \int_0^{\infty}
\frac{E\left\{f(\bm{X}_{i}^{(t)},\bm{X}_{m}^{(c)},\bm{\delta}_{i}^{(t)},\bm{\delta}_{m}^{(c)})\mid\bm{X}_{i}^{(t)},\bm{\delta}_{i}^{(t)}
\right\}I[g_{2}(\bm{X}_{i}^{(t)})>s]}{G^{(t)}[g_{1}(\bm{X}_{i}^{(t)})]G^{(c)}[g_{2}(\bm{X}_{i}^{(t)})]}
\cdot\frac{dM_j^{G^{(c)}}(s)}{y^{(c)}(s)}.
\end{eqnarray*}

{\section{An Example When $L=3$}

In this section, we provide an example for the explicit form of $K_{ij}^A$ and $L_{ij}^A$ with their estimators in the presence of three endpoints.
As discussed in Section 2.2 of the main manuscript, the estimator for $\pi_t$ takes the form,
$$
\widehat{\pi}_t  = \sum_{l=1}^L \widehat{\pi}_{tl} ,
$$
where
\begin{eqnarray*}
\widehat{\pi}_{t1} &=&\widehat{P}\{T_1^{(t)}\wedge \tau > T_1^{(c)}\wedge \tau + \zeta_1\} = \frac{1}{N_tN_c}\sum_{i=1}^{N_t}\sum_{j=1}^{N_c}\frac{I(X_{1,i}^{(t)}>X_{1,j}^{(c)}+\zeta_1)\delta_{1,j}^{(c)}}{\widehat{G}^{(t)}(X_{1,j}^{(c)}+\zeta_1)\widehat{G}^{(c)}(X_{1,j}^{(c)})},\\
\widehat{\pi}_{t2}  &=&\widehat{P}\{T_2^{(t)}\wedge \tau > T_2^{(c)}\wedge \tau + \zeta_2,{\cal U}_1\} \\
&=&\frac{1}{N_tN_c}\sum_{i=1}^{N_t}\sum_{j=1}^{N_c}\frac{I(X_{2,i}^{(t)}>X_{2,j}^{(c)}+\zeta_2,X_{1,i}^{(t)}> X_{1,j}^{(c)}-\zeta_1)\delta_{2,j}^{(c)}\delta_{1,j}^{(c)}}{
\widehat{G}^{(t)}\{(X_{2,j}^{(c)}+\zeta_2)\vee (X_{1,j}^{(c)}-\zeta_1)\}
\widehat{G}^{(c)}(X_{2,j}^{(c)}\vee X_{1,j}^{(c)})} \\
&-&
\frac{1}{N_tN_c}\sum_{i=1}^{N_t}\sum_{j=1}^{N_c}\frac{I(X_{2,i}^{(t)}>X_{2,j}^{(c)}+\zeta_2,X_{1,i}^{(t)}> X_{1,j}^{(c)}+\zeta_1)\delta_{2,j}^{(c)}\delta_{1,j}^{(c)}}{\widehat{G}^{(t)}\{(X_{2,j}^{(c)}+\zeta_2)\vee (X_{1,j}^{(c)}+\zeta_1)\}\widehat{G}^{(c)}(X_{2,j}^{(c)}\vee X_{1,j}^{(c)})}\\
\widehat{\pi}_{t3} &=&\widehat{P}\{T_3^{(t)}\wedge \tau > T_3^{(c)}\wedge \tau + \zeta_3,\cap_{k=1}^{2}{{\cal U}_k}\} \\
&=&\frac{1}{N_tN_c}\sum_{i=1}^{N_t}\sum_{j=1}^{N_c}\frac{I(X_{3,i}^{(t)}>X_{3,j}^{(c)}+\zeta_3,X_{2,i}^{(t)}> X_{2,j}^{(c)}-\zeta_2,X_{1,i}^{(t)}> X_{1,j}^{(c)}-\zeta_1)\delta_{3,j}^{(c)}\delta_{2,j}^{(c)}\delta_{1,j}^{(c)}}{\widehat{G}^{(t)}\{(X_{3,j}^{(c)}+\zeta_3)\vee (X_{2,j}^{(c)}-\zeta_2) \vee (X_{1,j}^{(c)}-\zeta_1)\}\widehat{G}^{(c)}(X_{3,j}^{(c)}\vee X_{2,j}^{(c)} \vee X_{1,j}^{(c)})} \\
&-&
\frac{1}{N_tN_c}\sum_{i=1}^{N_t}\sum_{j=1}^{N_c}\frac{I(X_{3,i}^{(t)}>X_{3,j}^{(c)}+\zeta_3,X_{2,i}^{(t)}> X_{2,j}^{(c)}-\zeta_2,X_{1,i}^{(t)}> X_{1,j}^{(c)}+\zeta_1)\delta_{3,j}^{(c)}\delta_{2,j}^{(c)}\delta_{1,j}^{(c)}}{\widehat{G}^{(t)}\{(X_{3,j}^{(c)}+\zeta_3)\vee (X_{2,j}^{(c)}-\zeta_2)\vee (X_{1,j}^{(c)}+\zeta_1)\}\widehat{G}^{(c)}(X_{3,j}^{(c)}\vee X_{2,j}^{(c)} \vee X_{1,j}^{(c)})} \\
&-&
\frac{1}{N_tN_c}\sum_{i=1}^{N_t}\sum_{j=1}^{N_c}\frac{I(X_{3,i}^{(t)}>X_{3,j}^{(c)}+\zeta_3,X_{2,i}^{(t)}> X_{2,j}^{(c)}+\zeta_2,X_{1,i}^{(t)}> X_{1,j}^{(c)}-\zeta_1)\delta_{3,j}^{(c)}\delta_{2,j}^{(c)}\delta_{1,j}^{(c)}}{\widehat{G}^{(t)}\{(X_{3,j}^{(c)}+\zeta_3)\vee(X_{2,j}^{(c)}+\zeta_2)\vee (X_{1,j}^{(c)}-\zeta_1)\}\widehat{G}^{(c)}(X_{3,j}^{(c)}\vee X_{2,j}^{(c)} \vee X_{1,j}^{(c)})} \\
&+&
\frac{1}{N_tN_c}\sum_{i=1}^{N_t}\sum_{j=1}^{N_c}\frac{I(X_{3,i}^{(t)}>X_{3,j}^{(c)}+\zeta_3,X_{2,i}^{(t)}> X_{2,j}^{(c)}+\zeta_2,X_{1,i}^{(t)}> X_{1,j}^{(c)}+\zeta_1)\delta_{3,j}^{(c)}\delta_{2,j}^{(c)}\delta_{1,j}^{(c)}}{\widehat{G}^{(t)}\{(X_{3,j}^{(c)}+\zeta_3)\vee(X_{2,j}^{(c)}+\zeta_2)\vee(X_{1,j}^{(c)}+\zeta_1)\}\widehat{G}^{(c)}(X_{3,j}^{(c)}\vee X_{2,j}^{(c)} \vee X_{1,j}^{(c)})}.
\end{eqnarray*}
The above estimator can be expressed as the summation of 7 terms taking the form of
$$
\widehat{P}_0^{(q)} \doteq \frac{1}{N_tN_c}\sum_{i=1}^{N_t}\sum_{j=1}^{N_c}\widehat{P}_{ij}^{(q)}=\frac{1}{N_tN_c}\sum_{i=1}^{N_t}\sum_{j=1}^{N_c}\frac{f^{(q)}(\bm{X}_{i}^{(t)},\bm{X}_{j}^{(c)},\bm{\delta}_{i}^{(t)},\bm{\delta}_{j}^{(c)})}{\widehat{G}^{(t)}[g_1^{(q)}(\bm{X}_{j}^{(c)})]\widehat{G}^{(c)}[g_2^{(q)}(\bm{X}_{j}^{(c)})]}, q = 1,\dots,7,
$$
where $\bm{X}_{i}^{(t)}\doteq (X_{1,i}^{(t)},X_{2,i}^{(t)},X_{3,i}^{(t)})$, $\bm{X}_{j}^{(c)}\doteq (X_{1,j}^{(c)},X_{2,j}^{(c)},X_{3,j}^{(c)})$, 
$\bm{\delta}_{i}^{(t)}\doteq (\delta_{1,i}^{(t)},\delta_{2,i}^{(t)},\delta_{3,i}^{(t)})$ and $\bm{\delta}_{j}^{(c)}\doteq (\delta_{1,j}^{(c)},\delta_{2,j}^{(c)},\delta_{3,j}^{(c)})$.
Define
\begin{align*}
&f^{(1)}(\bm{X}_{i}^{(t)},\bm{X}_{j}^{(c)},\bm{\delta}_{i}^{(t)},\bm{\delta}_{j}^{(c)}) \doteq I(X_{1,i}^{(t)}>X_{1,j}^{(c)}+\zeta_1)\delta_{1,j}^{(c)};\\
&g_1^{(1)}(\bm{X}_{j}^{(c)}) \doteq X_{1,j}^{(c)}+\zeta_1; g_2^{(1)}(\bm{X}_{j}^{(c)}) \doteq X_{1,j}^{(c)};\\
&f^{(2)}(\bm{X}_{i}^{(t)},\bm{X}_{j}^{(c)},\bm{\delta}_{i}^{(t)},\bm{\delta}_{j}^{(c)}) \doteq I(X_{2,i}^{(t)}>X_{2,j}^{(c)}+\zeta_2,X_{1,i}^{(t)}> X_{1,j}^{(c)}-\zeta_1)\delta_{2,j}^{(c)}\delta_{1,j}^{(c)};\\
&g_1^{(2)}(\bm{X}_{j}^{(c)}) \doteq (X_{2,j}^{(c)}+\zeta_2)\vee (X_{1,j}^{(c)}-\zeta_1; g_2^{(2)}(\bm{X}_{j}^{(c)}) \doteq X_{2,j}^{(c)}\vee X_{1,j}^{(c)};\\
&f^{(3)}(\bm{X}_{i}^{(t)},\bm{X}_{j}^{(c)},\bm{\delta}_{i}^{(t)},\bm{\delta}_{j}^{(c)}) \doteq -I(X_{2,i}^{(t)}>X_{2,j}^{(c)}+\zeta_2,X_{1,i}^{(t)}> X_{1,j}^{(c)}+\zeta_1)\delta_{2,j}^{(c)}\delta_{1,j}^{(c)};\\
&g_1^{(3)}(\bm{X}_{j}^{(c)}) \doteq (X_{2,j}^{(c)}+\zeta_2)\vee (X_{1,j}^{(c)}+\zeta_1; g_2^{(3)}(\bm{X}_{j}^{(c)}) \doteq X_{2,j}^{(c)}\vee X_{1,j}^{(c)};
\end{align*}
\begin{align*}
&f^{(4)}(\bm{X}_{i}^{(t)},\bm{X}_{j}^{(c)},\bm{\delta}_{i}^{(t)},\bm{\delta}_{j}^{(c)}) \doteq I(X_{3,i}^{(t)}>X_{3,j}^{(c)}+\zeta_3,X_{2,i}^{(t)}> X_{2,j}^{(c)}-\zeta_2,X_{1,i}^{(t)}> X_{1,j}^{(c)}-\zeta_1)\delta_{3,j}^{(c)}\delta_{2,j}^{(c)}\delta_{1,j}^{(c)};\\
&g_1^{(4)}(\bm{X}_{j}^{(c)}) \doteq (X_{3,j}^{(c)}+\zeta_3)\vee (X_{2,j}^{(c)}-\zeta_2) \vee (X_{1,j}^{(c)}-\zeta_1; g_2^{(4)}(\bm{X}_{j}^{(c)}) \doteq X_{3,j}^{(c)}\vee X_{2,j}^{(c)} \vee X_{1,j}^{(c)};\\
&f^{(5)}(\bm{X}_{i}^{(t)},\bm{X}_{j}^{(c)},\bm{\delta}_{i}^{(t)},\bm{\delta}_{j}^{(c)}) \doteq -I(X_{3,i}^{(t)}>X_{3,j}^{(c)}+\zeta_3,X_{2,i}^{(t)}> X_{2,j}^{(c)}-\zeta_2,X_{1,i}^{(t)}> X_{1,j}^{(c)}+\zeta_1)\delta_{3,j}^{(c)}\delta_{2,j}^{(c)}\delta_{1,j}^{(c)};\\
&g_1^{(5)}(\bm{X}_{j}^{(c)}) \doteq (X_{3,j}^{(c)}+\zeta_3)\vee (X_{2,j}^{(c)}-\zeta_2) \vee (X_{1,j}^{(c)}+\zeta_1; g_2^{(5)}(\bm{X}_{j}^{(c)}) \doteq X_{3,j}^{(c)}\vee X_{2,j}^{(c)} \vee X_{1,j}^{(c)};\\
&f^{(6)}(\bm{X}_{i}^{(t)},\bm{X}_{j}^{(c)},\bm{\delta}_{i}^{(t)},\bm{\delta}_{j}^{(c)}) \doteq -I(X_{3,i}^{(t)}>X_{3,j}^{(c)}+\zeta_3,X_{2,i}^{(t)}> X_{2,j}^{(c)}+\zeta_2,X_{1,i}^{(t)}> X_{1,j}^{(c)}-\zeta_1)\delta_{3,j}^{(c)}\delta_{2,j}^{(c)}\delta_{1,j}^{(c)};\\
&g_1^{(6)}(\bm{X}_{j}^{(c)}) \doteq (X_{3,j}^{(c)}+\zeta_3)\vee (X_{2,j}^{(c)}+\zeta_2) \vee (X_{1,j}^{(c)}-\zeta_1; g_2^{(6)}(\bm{X}_{j}^{(c)}) \doteq X_{3,j}^{(c)}\vee X_{2,j}^{(c)} \vee X_{1,j}^{(c)};\\
&f^{(7)}(\bm{X}_{i}^{(t)},\bm{X}_{j}^{(c)},\bm{\delta}_{i}^{(t)},\bm{\delta}_{j}^{(c)}) \doteq I(X_{3,i}^{(t)}>X_{3,j}^{(c)}+\zeta_3,X_{2,i}^{(t)}> X_{2,j}^{(c)}+\zeta_2,X_{1,i}^{(t)}> X_{1,j}^{(c)}+\zeta_1)\delta_{3,j}^{(c)}\delta_{2,j}^{(c)}\delta_{1,j}^{(c)};\\
&g_1^{(7)}(\bm{X}_{j}^{(c)}) \doteq (X_{3,j}^{(c)}+\zeta_3)\vee (X_{2,j}^{(c)}+\zeta_2) \vee (X_{1,j}^{(c)}+\zeta_1; g_2^{(7)}(\bm{X}_{j}^{(c)}) \doteq X_{3,j}^{(c)}\vee X_{2,j}^{(c)} \vee X_{1,j}^{(c)};
\end{align*}
Following the derivation in Section S3, we can write
$$
\sqrt{N_t+N_c}\left(\widehat{\pi}_t - \pi_t\right)  = \sqrt{N_t+N_c} \sum_{q=1}^7\left(\widehat{P}_0^{(q)} - P_0^{(q)}\right) = \frac{\sqrt{N_t+N_c}}{N_tN_c}\sum_{i=1}^{N_t}\sum_{j=1}^{N_c}K_{ij}^{A} + o_p(1),
$$
where $K_{ij}^{A} \doteq \sum_{q=1}^7\xi_{ij}^{(q)}$ with
\begin{eqnarray*}
\xi_{ij}^{(q)} &\doteq& (P_{ij}^{(q)} - P_0^{(q)}) + \int_0^{\infty}\frac{E\left\{f^{(q)}(\bm{X}_{k}^{(t)},\bm{X}_{j}^{(c)},\bm{\delta}_{k}^{(t)},\bm{\delta}_{j}^{(c)})\mid \bm{X}_j^{(c)}, \bm{\delta}_j^{(c)}\right\}I[g_{1}^{(q)}(\bm{X}_{j}^{(c)})>s]}{G^{(t)}[g_{1}^{(q)}(\bm{X}_{j}^{(c)})]G^{(c)}[g_{2}^{(q)}(\bm{X}_{j}^{(c)})]}\cdot\frac{dM_i^{G^{(t)}}(s)}{y^{(t)}(s)}\\
&+& \int_0^{\infty}E\left\{\frac{f^{(q)}(\bm{X}_{i}^{(t)},\bm{X}_{m}^{(c)},\bm{\delta}_{i}^{(t)},\bm{\delta}_{m}^{(c)})I[g_{2}^{(q)}(\bm{X}_{m}^{(c)})>s]}{G^{(t)}[g_{1}^{(q)}(\bm{X}_{m}^{(c)})]G^{(c)}[g_{2}^{(q)}(\bm{X}_{m}^{(c)})]} \biggm| \bm{X}_i^{(t)}, \bm{\delta}_i^{(t)}\right\}\cdot\frac{dM_j^{G^{(c)}}(s)}{y^{(c)}(s)}.
\end{eqnarray*}
Similarly, we can write
$$
\sqrt{N_t+N_c}\left(\widehat{\pi}_c - \pi_c\right)  = \frac{\sqrt{N_t+N_c}}{N_tN_c}\sum_{i=1}^{N_t}\sum_{j=1}^{N_c}L_{ij}^{A} + o_p(1),
$$
where $L_{ij}^{A} \doteq \sum_{q^*=1}^7\eta_{ij}^{(q^*)}$ with
\begin{eqnarray*}
\eta_{ij}^{(q^*)} &\doteq& (P_{ij}^{(q^*)} - P_0^{(q^*)}) + \int_0^{\infty}E\left\{\frac{f^{(q^*)}(\bm{X}_{k}^{(t)},\bm{X}_{j}^{(c)},\bm{\delta}_{k}^{(t)},\bm{\delta}_{j}^{(c)})I[g_{1}^{(q^*)}(\bm{X}_{k}^{(t)})>s]}{G^{(t)}[g_{1}^{(q^*)}(\bm{X}_{k}^{(t)})]G^{(c)}[g_{2}^{(q^*)}(\bm{X}_{k}^{(t)})]}
\biggm| \bm{X}_{j}^{(t)},\bm{\delta}_{j}^{(t)}\right\}
\cdot\frac{dM_i^{G^{(t)}}(s)}{y^{(t)}(s)}\\
&+& \int_0^{\infty}
\frac{E\left\{f^{(q^*)}(\bm{X}_{i}^{(t)},\bm{X}_{m}^{(c)},\bm{\delta}_{i}^{(t)},\bm{\delta}_{m}^{(c)})\mid\bm{X}_{i}^{(t)},\bm{\delta}_{i}^{(t)}
\right\}I[g_{2}^{(q^*)}(\bm{X}_{i}^{(t)})>s]}{G^{(t)}[g_{1}^{(q^*)}(\bm{X}_{i}^{(t)})]G^{(c)}[g_{2}^{(q^*)}(\bm{X}_{i}^{(t)})]}
\cdot\frac{dM_j^{G^{(c)}}(s)}{y^{(c)}(s)}
\end{eqnarray*}
for $f^{(q^*)}, g_{1}^{(q^*)}$ and $g_{2}^{(q^*)}, q^* = 1,\cdots, 7$ corresponding to $\widehat{\pi}_c$.
}

{
\section{Additional Simulation Studies}

\subsection{Simulation Studies to Investigate Settings Based on Different Distributional Assumption}

To further evaluate the robustness of our proposed method under different survival distributions, we conducted additional simulation studies using time to event outcome generated from Weibull distribution with a shape parameter of 2. We considered two simulation settings, following the same data generation steps for the first two scenarios based on exponential distribution. The only modification involved replacing the exponential distribution with a Weibull distribution to generate survival and censoring times. Specifically, the cumulative distribution functions $\{F_{l(t)}(\cdot), F_{l(c)}(\cdot), l=1, 2, 3\}$ and $F_C(\cdot)$ correspond to Weibull distributions with a shape parameter of 2 and scale parameters $\{1/\lambda_l^{(t)}, 1/\lambda_l^{(c)}, l=1, 2, 3\}$ and $1/\lambda_{C}$, respectively. The results are summarized in Tables \ref{tab:S1}-\ref{tab:S5}, which show that our proposed method performs consistently well under these new settings.

\begin{table}[H]
  \centering 
  \caption{Simulation results on bias (BIAS) in estimating the win probability for treatment and control ($\pi_t$ and $\pi_c$, respectively), and bias (BIAS) in estimating win ratio, average analytical standard error estimate of $\log(\widehat{WR})$ (ASE), empirical standard error of $\log(\widehat{WR})$ (ESE) and empirical coverage probability (CP) of 95\% CIs for win ratio across 5000 replicates with $\zeta = 0$ and $\tau\in \{18, 36\}$ in settings of Weibull distributions.}
\begin{tabular}{c|c|c|c|r|c|r|c|c|c|c|c}
\hline
& & & \multicolumn{2}{c|}{$\pi_t$} & \multicolumn{2}{c|}{$\pi_c$} & \multicolumn{5}{c}{WR} \\
\cline{4-12}  Setting  & $\tau$ & $(N_t,N_c)$ & TRUE   & BIAS  & TRUE   & BIAS  & TRUE   & BIAS  & ASE   & ESE   & CP \\
\hline
I & 18 & (100,100) & 0.411 & -0.001 & 0.411 & $<$0.001 & 1.000 & 0.019 & 0.203 & 0.205 & 0.942 \\
&       & (200,200) & 0.411 & $<$0.001 & 0.447 & -0.001 & 1.000 & 0.013 & 0.143 & 0.145 & 0.944 \\
&       & (400,400) & 0.411 & -0.001 & 0.411 & $<$0.001 & 1.000 & 0.004 & 0.101 & 0.102 & 0.949 \\
& 36 & (100,100) & 0.499 & -0.005 & 0.499 & -0.004 & 1.000 & 0.017 & 0.199 & 0.201 & 0.947 \\
&       & (200,200) & 0.499 & -0.002 & 0.499 & -0.004 & 1.000 & 0.013 & 0.140  & 0.140  & 0.947 \\
&       & (400,400) & 0.499 & -0.002 & 0.499 & -0.002 & 1.000 & 0.003 & 0.099 & 0.099 & 0.947 \\
\hline
II & 18 & (100,100) & 0.520  & -0.001 & 0.350  & $<$0.001 & 1.487 & 0.033 & 0.196 & 0.199 & 0.949 \\
&       & (200,200) & 0.520  & 0.001 & 0.350  & -0.001 & 1.487 & 0.025 & 0.139 & 0.140  & 0.949 \\
&       & (400,400) & 0.520  & $<$0.001 & 0.350  & $<$0.001 & 1.487 & 0.009 & 0.098 & 0.100 & 0.947 \\
& 36 & (100,100) & 0.641 & -0.005 & 0.358 & -0.003 & 1.790 & 0.048 & 0.200 & 0.199 & 0.954 \\
&       & (200,200) & 0.641 & -0.003 & 0.358 & -0.003 & 1.791 & 0.032 & 0.142 & 0.140  & 0.956 \\ 
&       & (400,400) & 0.641 & -0.002 & 0.358 & -0.002 & 1.791 & 0.016 & 0.100 & 0.099 & 0.954 \\
\hline
\end{tabular}
\label{tab:S1}
\end{table}

\begin{table}[H]
  \centering 
  \caption{Simulation results on bias (BIAS) in estimating the win probability for treatment and control ($\pi_t^{(\text{no})}$ and $\pi_c^{(\text{no})}$, respectively), and bias (BIAS) in estimating win ratio, average analytical standard error estimate of $\log(\widehat{WR}^{(\text{no})})$ (ASE), empirical standard error of $\log(\widehat{WR}^{(\text{no})})$ (ESE) and empirical coverage probability (CP) of 95\% CIs for the naive win ratio estimator without IPCW adjustment across 5000 replicates with $\zeta = 0$ and $\tau\in \{18, 36\}$ in settings of Weibull distributions.}
\begin{tabular}{c|c|c|c|r|c|r|c|c|c|c|c}
\hline
& & & \multicolumn{2}{c|}{$\pi_t^{(\text{no})}$} & \multicolumn{2}{c|}{$\pi_c^{(\text{no})}$} & \multicolumn{5}{c}{WR$^{(\text{no})}$} \\
\cline{4-12}  Setting  & $\tau$ & $(N_t,N_c)$ & TRUE   & BIAS  & TRUE   & BIAS  & TRUE   & BIAS  & ASE   & ESE   & CP \\
\hline
I & 18 & (100,100) & 0.411 & -0.037 & 0.411 & -0.035 & 1.000 & 0.017 & 0.197 & 0.200 & 0.944 \\
&       & (200,200) & 0.411 & -0.035 & 0.411 & -0.036 & 1.000 & 0.011 & 0.139 & 0.140 & 0.945 \\ 
&       & (400,400) & 0.411 & -0.036 & 0.411 & -0.036 & 1.000 & 0.004 & 0.098 & 0.099 & 0.944 \\ 
& 36 & (100,100) & 0.499 & -0.065 & 0.499 & -0.064 & 1.000 & 0.011 & 0.169 & 0.172 & 0.946 \\
&       & (200,200) & 0.499 & -0.064 & 0.499 & -0.065 & 1.000 & 0.010 & 0.119 & 0.120 & 0.946 \\
&       & (400,400) & 0.499 & -0.065 & 0.499 & -0.064 & 1.000 & 0.003 & 0.085 & 0.086 & 0.944 \\
\hline
II & 18 & (100,100) & 0.520  & -0.044 & 0.349 & -0.027 & 1.487 & 0.024 & 0.190 & 0.193 & 0.948 \\ 
&       & (200,200) & 0.520  & -0.042 & 0.350 & -0.028 & 1.487 & 0.016 & 0.135 & 0.136 & 0.952 \\
&       & (400,400) & 0.520  & -0.043 & 0.350 & -0.028 & 1.487 & 0.001 & 0.095 & 0.097 & 0.947 \\ 
& 36 & (100,100) & 0.641 & -0.087 & 0.358 & -0.025 & 1.791 & -0.097 & 0.170 & 0.171 & 0.928 \\
&       & (200,200) & 0.641 & -0.086 & 0.358 & -0.026 & 1.791 & -0.103 & 0.121 & 0.121 & 0.909 \\ 
&       & (400,400) & 0.641 & -0.086 & 0.358 & -0.025 & 1.791 & -0.117 & 0.085 & 0.086 & 0.862 \\
\hline
\end{tabular}
\label{tab:S2}
\end{table}

\begin{table}[H]
  \centering 
  \caption{ Simulation results on bias (BIAS) in estimating the win probability for treatment and control ($\pi_t$ and $\pi_c$, respectively), and bias (BIAS) in estimating win ratio, average analytical standard error estimate of $\log(\widehat{WR})$ (ASE), empirical standard error of $\log(\widehat{WR})$ (ESE) and empirical coverage probability (CP) of 95\% CIs for win ratio across 5000 replicates with different combinations of $\tau$s and $\zeta$s in different settings of Weibull distributions ($N_t=N_c=200$).}
    \begin{tabular}{c|c|r|c|r|c|r|c|c|c|c|c}
    \hline
    \multirow{2}{*}{} & &  & \multicolumn{2}{c|}{$\pi_t$} & \multicolumn{2}{c|}{$\pi_c$} & \multicolumn{5}{c}{WR} \\
\cline{4-12}      Setting    & $\tau$ &   $\zeta$    & TRUE   & BIAS  & TRUE   & BIAS  & TRUE   & BIAS  & ASE   & ESE   & CP \\
\hline
I& 18    & 0     & 0.411 & $<$0.001 & 0.411 & -0.001 & 1.000 & 0.013 & 0.143 & 0.145 & 0.944 \\
     & 18    & 2    & 0.350  & -0.001 & 0.350 & -0.002 & 1.000 & 0.015 & 0.161 & 0.163 & 0.949 \\
     & 18    & 4    & 0.285 & -0.001 & 0.285 & -0.001 & 1.000 & 0.020 & 0.182 & 0.185 & 0.940 \\
     & 18    & 6    & 0.220  & -0.001 & 0.220 & -0.001 & 1.000 & 0.025 & 0.208 & 0.210 & 0.943 \\
     & 36    & 0    & 0.499 & -0.002 & 0.499 & -0.004 & 1.000 & 0.013 & 0.140 & 0.140 & 0.947 \\
     & 36    & 2    & 0.480  & -0.002 & 0.480 & -0.004 & 1.000 & 0.014 & 0.144 & 0.145 & 0.946 \\
     & 36    & 4    & 0.453 & -0.002 & 0.453 & -0.004 & 1.000 & 0.014 & 0.149 & 0.149 & 0.943 \\
     & 36    & 6    & 0.421 & -0.002 & 0.421 & -0.003 & 1.000 & 0.015 & 0.155 & 0.155 & 0.945 \\
\hline
II & 18    & 0    & 0.520  & 0.001 & 0.350 & -0.001 & 1.487 & 0.025 & 0.139 & 0.140 & 0.949 \\
     & 18    & 2   & 0.453 & -0.001 & 0.303 & -0.002 & 1.497 & 0.030 & 0.154 & 0.156 & 0.947 \\
     & 18    & 4   & 0.378 & $<$0.001 & 0.249 & -0.002 & 1.515 & 0.039 & 0.173 & 0.175 & 0.950 \\
     & 18    & 6   & 0.298 & $<$0.001 & 0.193 & -0.001 & 1.545 & 0.048 & 0.197 & 0.199 & 0.949 \\
     & 36    & 0   & 0.641 & -0.003 & 0.358 & -0.003 & 1.791 & 0.032 & 0.142 & 0.140 & 0.956 \\
     & 36    & 2   & 0.626 & -0.003 & 0.349 & -0.003 & 1.796 & 0.030 & 0.143 & 0.142 & 0.956 \\
     & 36    & 4   & 0.604 & -0.003 & 0.334 & -0.003 & 1.808 & 0.029 & 0.146 & 0.145 & 0.957 \\
     & 36    & 6   & 0.573 & -0.003 & 0.313 & -0.003 & 1.830 & 0.032 & 0.151 & 0.150 & 0.955 \\
\hline
    \end{tabular}
  \label{tab:S3}
\end{table}

\begin{table}[H]
  \centering 
  \caption{ Simulation results on bias (BIAS) in estimating the win probability for treatment and control ($\pi_t^{(\text{no})}$ and $\pi_c^{(\text{no})}$, respectively), and bias (BIAS) in estimating win ratio, average analytical standard error estimate of $\log(\widehat{WR}^{(\text{no})})$ (ASE), empirical standard error of $\log(\widehat{WR}^{(\text{no})})$ (ESE) and empirical coverage probability (CP) of 95\% CIs for the naive win ratio estimator without IPCW adjustment across 5000 replicates with different combinations of $\tau$s and $\zeta$s in different settings of Weibull distributions ($N_t=N_c=200$).}
    \begin{tabular}{c|c|r|c|r|c|r|c|c|c|c|c}
    \hline
    \multirow{2}{*}{} & &  & \multicolumn{2}{c|}{$\pi_t^{(\text{no})}$} & \multicolumn{2}{c|}{$\pi_c^{(\text{no})}$} & \multicolumn{5}{c}{WR$^{(\text{no})}$} \\
\cline{4-12}      Setting    & $\tau$ &   $\zeta$    & TRUE   & BIAS  & TRUE   & BIAS  & TRUE   & BIAS  & ASE   & ESE   & CP \\
\hline
I& 18    & 0     & 0.411 & -0.035 & 0.411 & -0.036 & 1.000 & 0.011 & 0.139 & 0.140 & 0.945 \\ 
     & 18    & 2    & 0.350 & -0.030 & 0.350 & -0.031 & 1.000 & 0.014 & 0.156 & 0.158 & 0.948 \\
     & 18    & 4    & 0.285 & -0.025 & 0.285 & -0.026 & 1.000 & 0.019 & 0.177 & 0.180 & 0.942 \\
     & 18    & 6    & 0.220 & -0.020 & 0.220 & -0.020 & 1.000 & 0.023 & 0.202 & 0.203 & 0.941 \\ 
     & 36    & 0    & 0.499 & -0.064 & 0.499 & -0.065 & 1.000 & 0.010 & 0.119 & 0.120 & 0.946 \\
     & 36    & 2    & 0.480 & -0.073 & 0.480 & -0.074 & 1.000 & 0.012 & 0.126 & 0.127 & 0.948 \\
     & 36    & 4    & 0.453 & -0.079 & 0.453 & -0.081 & 1.000 & 0.012 & 0.133 & 0.134 & 0.947 \\ 
     & 36    & 6    & 0.421 & -0.083 & 0.421 & -0.084 & 1.000 & 0.013 & 0.141 & 0.142 & 0.948 \\ 
\hline
II & 18    & 0    & 0.520 & -0.042 & 0.35 & -0.028 & 1.487 & 0.016 & 0.135 & 0.136 & 0.952 \\ 
     & 18    & 2   & 0.453 & -0.038 & 0.303 & -0.026 & 1.497 & 0.022 & 0.150 & 0.152 & 0.950 \\
     & 18    & 4   & 0.378 & -0.032 & 0.249 & -0.022 & 1.515 & 0.032 & 0.169 & 0.171 & 0.949 \\ 
     & 18    & 6   & 0.298 & -0.026 & 0.193 & -0.018 & 1.545 & 0.041 & 0.192 & 0.194 & 0.949 \\ 
     & 36    & 0   & 0.641 & -0.086 & 0.358 & -0.026 & 1.791 & -0.103 & 0.121 & 0.121 & 0.909 \\
     & 36    & 2   & 0.626 & -0.098 & 0.349 & -0.037 & 1.796 & -0.087 & 0.126 & 0.126 & 0.925 \\
     & 36    & 4   & 0.604 & -0.108 & 0.334 & -0.046 & 1.808 & -0.070 & 0.132 & 0.132 & 0.934 \\
     & 36    & 6   & 0.573 & -0.115 & 0.313 & -0.053 & 1.830 & -0.052 & 0.139 & 0.139 & 0.943 \\
\hline
    \end{tabular}
  \label{tab:S4}
\end{table}

\begin{table}[H]
  \centering \small
  \caption{ Simulation results for the empirical type I error rate and empirical power of the test based on win ratio estimator with IPCW adjustment (WR), naive win ratio estimator without IPCW adjustment (WR$^{(\text{no})}$) and log-rank test for comparing the time to the first occurred event (Logrank) across 5000 replicates with different combinations of $\tau$s and $\zeta$s in settings of Weibull distributions.}
    \begin{tabular}{c|c|c|c|c|c|c|c|c|c|c|c}
    \hline
    & & & \multicolumn{3}{c|}{$\zeta=0$} & \multicolumn{2}{c|}{$\zeta=2$} & \multicolumn{2}{c|}{$\zeta=4$} & \multicolumn{2}{c}{$\zeta=6$} \\
\cline{4-12} Setting   &   $\tau$    &    $(N_t,N_c)$ & WR  & WR$^{(\text{no})}$  & Logrank & WR  & WR$^{(\text{no})}$  & WR  & WR$^{(\text{no})}$  & WR & WR$^{(\text{no})}$\\
    \hline
    I & 18    & (100,100) & 0.052 & 0.049 & 0.054 & 0.047 & 0.051 & 0.046 & 0.050 & 0.047 & 0.051 \\
    & 18    & (200,200) & 0.054 & 0.051 & 0.052 & 0.051 & 0.050 & 0.056 & 0.053 & 0.052 & 0.052 \\ 
    & 18    & (400,400) & 0.047 & 0.051 & 0.054 & 0.048 & 0.052 & 0.050 & 0.048 & 0.050 & 0.048 \\ 
    & 36    & (100,100) & 0.049 & 0.047 & 0.054 & 0.046 & 0.045 & 0.045 & 0.045 & 0.047 & 0.046 \\
    & 36    & (200,200) & 0.054 & 0.050 & 0.049 & 0.057 & 0.053 & 0.055 & 0.050 & 0.051 & 0.049 \\
    & 36    & (400,400) & 0.051 & 0.051 & 0.052 & 0.052 & 0.050 & 0.051 & 0.052 & 0.054 & 0.051 \\ 
    \hline
    II & 18    & (100,100) & 0.644 & 0.654 & 0.383 & 0.583 & 0.598 & 0.526 & 0.543 & 0.481 & 0.487 \\ 
    & 18    & (200,200) & 0.892 & 0.908 & 0.669 & 0.844 & 0.856 & 0.789 & 0.802 & 0.731 & 0.740 \\
    & 18    & (400,400) & 0.991 & 0.994 & 0.913 & 0.982 & 0.985 & 0.961 & 0.967 & 0.936 & 0.944 \\
    & 36    & (100,100) & 0.911 & 0.917 & 0.502 & 0.904 & 0.907 & 0.898 & 0.899 & 0.891 & 0.888 \\
    & 36    & (200,200) & 0.996 & 0.996 & 0.797 & 0.995 & 0.995 & 0.994 & 0.993 & 0.994 & 0.993 \\ 
    & 36    & (400,400) & 1.000& 1.000 & 0.975 & 1.000 & 1.000 & 1.000 & 1.000 & 1.000 & 1.000 \\
    \hline
    \end{tabular}
  \label{tab:S5}
\end{table}

\subsection{Simulation Studies to Evaluate Impact of Inducing Common Censoring}

We have conducted simulation studies to investigate the impact of inducing common censoring when there exist different censoring mechanisms for different endpoints.
In this simulation study, we focus on the case with two time to event outcomes, $T_1$ and $T_2$, which are subject to right censoring by $C_1$ and $C_2,$ respectively. We have investigated three different methods to calculate the win statistic:
\begin{enumerate}
\item The proposed method with a induced common censoring time, $C_1\wedge C_2,$ whose survival function is estimated by Kaplan-Meier estimator and used in IPCW; denoted by $\text{WR}$;
\item The proposed method with a induced common censoring time, $C_1\wedge C_2,$ whose true survival function is used in IPCW; denoted by $\text{WR}^{(TC)}$;
\item The proposed method with the true joint censoring distribution for $(C_1, C_2)$ used in IPCW; denoted by $\text{WR}^{(TJ)}$;
\end{enumerate}
The estimator of these three methods differs only with the kernel function.  We present in two cases:
\begin{enumerate}
\item  $\zeta= 0$
\begin{enumerate}
\item For the first method (our proposal), the estimator for WR is
$$
\text{WR} = \frac{\widehat{\pi}_{t}}{\widehat{\pi}_{c}},
$$
where 
\begin{align*}
\widehat{\pi}_{t} =& \widehat{\pi}_{t1} + \widehat{\pi}_{t2}  \\
=& \frac{1}{N_tN_c}\sum_{i=1}^{N_t}\sum_{j=1}^{N_c}\frac{I(X_{1,i}^{(t)}>X_{1,j}^{(c)})\delta_{1,j}^{(c)}}{\widehat{G}^{(t)}(X_{1,j}^{(c)})\widehat{G}^{(c)}(X_{1,j}^{(c)})}\\
+&\frac{1}{N_tN_c}\sum_{i=1}^{N_t}\sum_{j=1}^{N_c}\frac{I(X_{2,i}^{(t)}>X_{2,j}^{(c)},X_{1,i}^{(t)}= X_{1,j}^{(c)}=\tau)\delta_{2,j}^{(c)}}{
\widehat{G}^{(t)}(\tau)
\widehat{G}^{(c)}(\tau)}
\end{align*}
and 
\begin{align*}
\widehat{\pi}_{c} =& \widehat{\pi}_{c1} + \widehat{\pi}_{c2}  \\
=& \frac{1}{N_tN_c}\sum_{i=1}^{N_t}\sum_{j=1}^{N_c}\frac{I(X_{1,i}^{(c)}>X_{1,j}^{(t)})\delta_{1,j}^{(t)}}{\widehat{G}^{(c)}(X_{1,j}^{(t)})\widehat{G}^{(t)}(X_{1,j}^{(t)})}\\
+&\frac{1}{N_tN_c}\sum_{i=1}^{N_t}\sum_{j=1}^{N_c}\frac{I(X_{2,i}^{(c)}>X_{2,j}^{(t)},X_{1,i}^{(c)}= X_{1,j}^{(t)}=\tau)\delta_{2,j}^{(t)}}{
\widehat{G}^{(c)}(\tau)
\widehat{G}^{(t)}(\tau)}
\end{align*}
\item For the second method, the estimator for WR is
$$
\text{WR}^{(TC)} = \frac{\widehat{\pi}_{t}^{(TC)}}{\widehat{\pi}_{c}^{(TC)}},
$$
where 
\begin{align*}
\widehat{\pi}_{t}^{(TC)} =& \frac{1}{N_tN_c}\sum_{i=1}^{N_t}\sum_{j=1}^{N_c}\frac{I(X_{1,i}^{(t)}>X_{1,j}^{(c)})\delta_{1,j}^{(c)}}{G^{(t)}(X_{1,j}^{(c)})G^{(c)}(X_{1,j}^{(c)})}\\
+&\frac{1}{N_tN_c}\sum_{i=1}^{N_t}\sum_{j=1}^{N_c}\frac{I(X_{2,i}^{(t)}>X_{2,j}^{(c)},X_{1,i}^{(t)}= X_{1,j}^{(c)}=\tau)\delta_{2,j}^{(c)}}{
G^{(t)}(\tau)
G^{(c)}(\tau)}
\end{align*}
and 
\begin{align*}
\widehat{\pi}_{c}^{(TC)} =& \frac{1}{N_tN_c}\sum_{i=1}^{N_t}\sum_{j=1}^{N_c}\frac{I(X_{1,i}^{(c)}>X_{1,j}^{(t)})\delta_{1,j}^{(t)}}{G^{(c)}(X_{1,j}^{(t)})G^{(t)}(X_{1,j}^{(t)})}\\
+&\frac{1}{N_tN_c}\sum_{i=1}^{N_t}\sum_{j=1}^{N_c}\frac{I(X_{2,i}^{(c)}>X_{2,j}^{(t)},X_{1,i}^{(c)}= X_{1,j}^{(t)}=\tau)\delta_{2,j}^{(t)}}{
G^{(c)}(\tau)
G^{(t)}(\tau)}
\end{align*}
\item For the third method, denote $S_{12}^{(\cdot)}$ as the joint survival distribution of $T_1$ and $T_2$, and the estimator for WR is
$$
\text{WR}^{(TJ)} = \frac{\widehat{\pi}_{t}^{(TJ)}}{\widehat{\pi}_{c}^{(TJ)}},
$$
where 
\begin{align*}
\widehat{\pi}_{t}^{(TJ)} =& \frac{1}{N_tN_c}\sum_{i=1}^{N_t}\sum_{j=1}^{N_c}\frac{I(X_{1,i}^{(t)}>X_{1,j}^{(c)})\delta_{1,j}^{(c)}}{S_{12}^{(t)}(X_{1,j}^{(c)},0)S_{12}^{(c)}(X_{1,j}^{(c)},0)}\\
+&\frac{1}{N_tN_c}\sum_{i=1}^{N_t}\sum_{j=1}^{N_c}\frac{I(X_{2,i}^{(t)}>X_{2,j}^{(c)},X_{1,i}^{(t)}= X_{1,j}^{(c)}=\tau)\delta_{2,j}^{(c)}}{
S_{12}^{(t)}\{\tau, X_{2,j}^{(c)}\}
S_{12}^{(c)}(\tau, X_{2,j}^{(c)})} 
\end{align*}
and 
\begin{align*}
\widehat{\pi}_{c}^{(TJ)} =& \frac{1}{N_tN_c}\sum_{i=1}^{N_t}\sum_{j=1}^{N_c}\frac{I(X_{1,i}^{(c)}>X_{1,j}^{(t)})\delta_{1,j}^{(t)}}{S_{12}^{(c)}(X_{1,j}^{(t)},0)S_{12}^{(t)}(X_{1,j}^{(t)},0)}\\
+&\frac{1}{N_tN_c}\sum_{i=1}^{N_t}\sum_{j=1}^{N_c}\frac{I(X_{2,i}^{(c)}>X_{2,j}^{(t)},X_{1,i}^{(c)}= X_{1,j}^{(t)}=\tau)\delta_{2,j}^{(t)}}{
S_{12}^{(c)}\{\tau, X_{2,j}^{(t)}\}
S_{12}^{(t)}(\tau, X_{2,j}^{(t)})} 
\end{align*}
\end{enumerate}
\item $\zeta> 0$
\begin{enumerate}
\item For the first method, the estimator for WR is
$$
\text{WR} = \frac{\widehat{\pi}_{t}}{\widehat{\pi}_{c}},
$$
where 
\begin{align*}
\widehat{\pi}_{t} =& \frac{1}{N_tN_c}\sum_{i=1}^{N_t}\sum_{j=1}^{N_c}\frac{I(X_{1,i}^{(t)}>X_{1,j}^{(c)}+\zeta_1)\delta_{1,j}^{(c)}}{\widehat{G}^{(t)}(X_{1,j}^{(c)}+\zeta_1)\widehat{G}^{(c)}(X_{1,j}^{(c)})}\\
+&\frac{1}{N_tN_c}\sum_{i=1}^{N_t}\sum_{j=1}^{N_c}\frac{I(X_{2,i}^{(t)}>X_{2,j}^{(c)}+\zeta_2,X_{1,i}^{(t)}> X_{1,j}^{(c)}-\zeta_1)\delta_{2,j}^{(c)}\delta_{1,j}^{(c)}}{
\widehat{G}^{(t)}\{(X_{2,j}^{(c)}+\zeta_2)\vee (X_{1,j}^{(c)}-\zeta_1)\}
\widehat{G}^{(c)}(X_{2,j}^{(c)}\vee X_{1,j}^{(c)})} \\
-&
\frac{1}{N_tN_c}\sum_{i=1}^{N_t}\sum_{j=1}^{N_c}\frac{I(X_{2,i}^{(t)}>X_{2,j}^{(c)}+\zeta_2,X_{1,i}^{(t)}> X_{1,j}^{(c)}+\zeta_1)\delta_{2,j}^{(c)}\delta_{1,j}^{(c)}}{\widehat{G}^{(t)}\{(X_{2,j}^{(c)}+\zeta_2)\vee (X_{1,j}^{(c)}+\zeta_1)\}\widehat{G}^{(c)}(X_{2,j}^{(c)}\vee X_{1,j}^{(c)})}
\end{align*}
and 
\begin{align*}
\widehat{\pi}_{c} =& \frac{1}{N_tN_c}\sum_{i=1}^{N_t}\sum_{j=1}^{N_c}\frac{I(X_{1,i}^{(c)}>X_{1,j}^{(t)}+\zeta_1)\delta_{1,j}^{(t)}}{\widehat{G}^{(c)}(X_{1,j}^{(t)}+\zeta_1)\widehat{G}^{(t)}(X_{1,j}^{(t)})}\\
+&\frac{1}{N_tN_c}\sum_{i=1}^{N_t}\sum_{j=1}^{N_c}\frac{I(X_{2,i}^{(c)}>X_{2,j}^{(t)}+\zeta_2,X_{1,i}^{(c)}> X_{1,j}^{(t)}-\zeta_1)\delta_{2,j}^{(t)}\delta_{1,j}^{(t)}}{
\widehat{G}^{(c)}\{(X_{2,j}^{(t)}+\zeta_2)\vee (X_{1,j}^{(t)}-\zeta_1)\}
\widehat{G}^{(t)}(X_{2,j}^{(t)}\vee X_{1,j}^{(t)})} \\
-&
\frac{1}{N_tN_c}\sum_{i=1}^{N_t}\sum_{j=1}^{N_c}\frac{I(X_{2,i}^{(c)}>X_{2,j}^{(t)}+\zeta_2,X_{1,i}^{(c)}> X_{1,j}^{(t)}+\zeta_1)\delta_{2,j}^{(t)}\delta_{1,j}^{(t)}}{\widehat{G}^{(c)}\{(X_{2,j}^{(t)}+\zeta_2)\vee (X_{1,j}^{(t)}+\zeta_1)\}\widehat{G}^{(t)}(X_{2,j}^{(t)}\vee X_{1,j}^{(t)})}
\end{align*}
\item For the second method, the estimator of the WR is
$$
\text{WR}^{(TC)} = \frac{\widehat{\pi}_{t}^{(TC)}}{\widehat{\pi}_{c}^{(TC)}},
$$
where 
\begin{align*}
\widehat{\pi}_{t}^{(TC)} =& \frac{1}{N_tN_c}\sum_{i=1}^{N_t}\sum_{j=1}^{N_c}\frac{I(X_{1,i}^{(t)}>X_{1,j}^{(c)}+\zeta_1)\delta_{1,j}^{(c)}}{G^{(t)}(X_{1,j}^{(c)}+\zeta_1)G^{(c)}(X_{1,j}^{(c)})}\\
+&\frac{1}{N_tN_c}\sum_{i=1}^{N_t}\sum_{j=1}^{N_c}\frac{I(X_{2,i}^{(t)}>X_{2,j}^{(c)}+\zeta_2,X_{1,i}^{(t)}> X_{1,j}^{(c)}-\zeta_1)\delta_{2,j}^{(c)}\delta_{1,j}^{(c)}}{
G^{(t)}\{(X_{2,j}^{(c)}+\zeta_2)\vee (X_{1,j}^{(c)}-\zeta_1)\}
G^{(c)}(X_{2,j}^{(c)}\vee X_{1,j}^{(c)})} \\
-&
\frac{1}{N_tN_c}\sum_{i=1}^{N_t}\sum_{j=1}^{N_c}\frac{I(X_{2,i}^{(t)}>X_{2,j}^{(c)}+\zeta_2,X_{1,i}^{(t)}> X_{1,j}^{(c)}+\zeta_1)\delta_{2,j}^{(c)}\delta_{1,j}^{(c)}}{G^{(t)}\{(X_{2,j}^{(c)}+\zeta_2)\vee (X_{1,j}^{(c)}+\zeta_1)\}G^{(c)}(X_{2,j}^{(c)}\vee X_{1,j}^{(c)})}
\end{align*}
and 
\begin{align*}
\widehat{\pi}_{c}^{(TC)} =& \frac{1}{N_tN_c}\sum_{i=1}^{N_t}\sum_{j=1}^{N_c}\frac{I(X_{1,i}^{(c)}>X_{1,j}^{(t)}+\zeta_1)\delta_{1,j}^{(t)}}{G^{(c)}(X_{1,j}^{(t)}+\zeta_1)G^{(t)}(X_{1,j}^{(t)})}\\
+&\frac{1}{N_tN_c}\sum_{i=1}^{N_t}\sum_{j=1}^{N_c}\frac{I(X_{2,i}^{(c)}>X_{2,j}^{(t)}+\zeta_2,X_{1,i}^{(c)}> X_{1,j}^{(t)}-\zeta_1)\delta_{2,j}^{(t)}\delta_{1,j}^{(t)}}{
G^{(c)}\{(X_{2,j}^{(t)}+\zeta_2)\vee (X_{1,j}^{(t)}-\zeta_1)\}
G^{(t)}(X_{2,j}^{(t)}\vee X_{1,j}^{(t)})} \\
-&
\frac{1}{N_tN_c}\sum_{i=1}^{N_t}\sum_{j=1}^{N_c}\frac{I(X_{2,i}^{(c)}>X_{2,j}^{(t)}+\zeta_2,X_{1,i}^{(c)}> X_{1,j}^{(t)}+\zeta_1)\delta_{2,j}^{(t)}\delta_{1,j}^{(t)}}{G^{(c)}\{(X_{2,j}^{(t)}+\zeta_2)\vee (X_{1,j}^{(t)}+\zeta_1)\}G^{(t)}(X_{2,j}^{(t)}\vee X_{1,j}^{(t)})}
\end{align*}
\item For the third method, denote $S_{12}^{(\cdot)}$ as the joint survival distribution of $T_1$ and $T_2$, and the estimator of WR is
$$
\text{WR}^{(TJ)} = \frac{\widehat{\pi}_{t}^{(TJ)}}{\widehat{\pi}_{c}^{(TJ)}},
$$
where 
\begin{align*}
\widehat{\pi}_{t}^{(TJ)} =& \frac{1}{N_tN_c}\sum_{i=1}^{N_t}\sum_{j=1}^{N_c}\frac{I(X_{1,i}^{(t)}>X_{1,j}^{(c)}+\zeta_1)\delta_{1,j}^{(c)}}{S_{12}^{(t)}(X_{1,j}^{(c)}+\zeta_1,0)S_{12}^{(c)}(X_{1,j}^{(c)},0)}\\
+&\frac{1}{N_tN_c}\sum_{i=1}^{N_t}\sum_{j=1}^{N_c}\frac{I(X_{2,i}^{(t)}>X_{2,j}^{(c)}+\zeta_2,X_{1,i}^{(t)}> X_{1,j}^{(c)}-\zeta_1)\delta_{2,j}^{(c)}\delta_{1,j}^{(c)}}{
S_{12}^{(t)}\{(X_{1,j}^{(c)}-\zeta_1), (X_{2,j}^{(c)}+\zeta_2)\}
S_{12}^{(c)}(X_{1,j}^{(c)}, X_{2,j}^{(c)})} \\
-&
\frac{1}{N_tN_c}\sum_{i=1}^{N_t}\sum_{j=1}^{N_c}\frac{I(X_{2,i}^{(t)}>X_{2,j}^{(c)}+\zeta_2,X_{1,i}^{(t)}> X_{1,j}^{(c)}+\zeta_1)\delta_{2,j}^{(c)}\delta_{1,j}^{(c)}}{S_{12}^{(t)}\{(X_{1,j}^{(c)}+\zeta_1), (X_{2,j}^{(c)}+\zeta_2)\}S_{12}^{(c)}(X_{1,j}^{(c)}, X_{2,j}^{(c)})}
\end{align*}
and 
\begin{align*}
\widehat{\pi}_{c}^{(TJ)} =& \frac{1}{N_tN_c}\sum_{i=1}^{N_t}\sum_{j=1}^{N_c}\frac{I(X_{1,i}^{(c)}>X_{1,j}^{(t)}+\zeta_1)\delta_{1,j}^{(t)}}{S_{12}^{(c)}(X_{1,j}^{(t)}+\zeta_1,0)S_{12}^{(t)}(X_{1,j}^{(t)},0)}\\
+&\frac{1}{N_tN_c}\sum_{i=1}^{N_t}\sum_{j=1}^{N_c}\frac{I(X_{2,i}^{(c)}>X_{2,j}^{(t)}+\zeta_2,X_{1,i}^{(c)}> X_{1,j}^{(t)}-\zeta_1)\delta_{2,j}^{(t)}\delta_{1,j}^{(t)}}{
S_{12}^{(c)}\{(X_{1,j}^{(t)}-\zeta_1),(X_{2,j}^{(t)}+\zeta_2)\}
S_{12}^{(t)}(X_{1,j}^{(t)}, X_{2,j}^{(t)})} \\
-&
\frac{1}{N_tN_c}\sum_{i=1}^{N_t}\sum_{j=1}^{N_c}\frac{I(X_{2,i}^{(c)}>X_{2,j}^{(t)}+\zeta_2,X_{1,i}^{(c)}> X_{1,j}^{(t)}+\zeta_1)\delta_{2,j}^{(t)}\delta_{1,j}^{(t)}}{S_{12}^{(c)}\{(X_{1,j}^{(t)}+\zeta_1), (X_{2,j}^{(t)}+\zeta_2)\}S_{12}^{(t)}(X_{1,j}^{(t)}, X_{2,j}^{(t)})}.
\end{align*}
\end{enumerate}
\end{enumerate}

The data are generated via the following steps:
\begin{enumerate}
\item  Generate $(T_1^{(\cdot)}, T_2^{(\cdot)})$ as 
$(F_{1(\cdot)}^{-1}\{\Phi(Z_1)\}, F_{2(\cdot)}^{-1}\{\Phi(Z_2)\}),$
where $F_{l(\cdot)}(\cdot)$ denotes the selected cumulative distribution function (CDF), $F_{l(\cdot)}^{-1}(\cdot)$ is the inverse function of $F_{l(\cdot)}(\cdot)$, $\Phi(\cdot)$ denotes the CDF of $N(0,1)$, and
$$
\left(\begin{array}{c} Z_1 \\ Z_2 \end{array}\right)\sim \text{Normal}
\begin{pmatrix}
\begin{pmatrix}
0\\
0
\end{pmatrix},
\begin{pmatrix}
1 & 0.5\\
0.5 & 1
\end{pmatrix}
\end{pmatrix}.
$$
\item Generate $(C_1^{(\cdot)}, C_2^{(\cdot)})$ as 
$(F_{C_1(\cdot)}^{-1}\{\Phi(Z_{C_1})\}, F_{C_2(\cdot)}^{-1}\{\Phi(Z_{C_2})\}),$
where $F_{C_l(\cdot)}(\cdot)$ denotes the selected cumulative distribution function (CDF), $F_{C_l(\cdot)}^{-1}(\cdot)$ is the inverse function of $F_{C_l(\cdot)}(\cdot)$, $\Phi(\cdot)$ denotes the CDF of $N(0,1)$, and
$$
\left(\begin{array}{c} Z_{C_1} \\ Z_{C_2} \end{array}\right)\sim \text{Normal}
\begin{pmatrix}
\begin{pmatrix}
0\\
0
\end{pmatrix},
\begin{pmatrix}
1 & \rho\\
\rho & 1
\end{pmatrix}
\end{pmatrix}.
$$
\end{enumerate}
When applying the first two methods, we induce a common censoring and observed data consist of realizations of 
$$ 
{X_l^{(\cdot)}} = \min\{T_l^{(\cdot)}\wedge \tau,C_1^{(\cdot)}\wedge C_2^{(\cdot)}\} 
\text{ and }
{\delta_l^{(\cdot)}} = I\{T_l^{(\cdot)}\wedge \tau \le C_1^{(\cdot)}\wedge C_2^{(\cdot)}\}, l=1, 2.
$$
When applying the third method, observed data consist of realizations of 
$$ 
{X_l^{(\cdot)}} = \min\{T_l^{(\cdot)}\wedge \tau,C_l^{(\cdot)}\} 
\text{ and }
{\delta_l^{(\cdot)}} = I\{T_l^{(\cdot)}\wedge \tau \le C_l^{(\cdot)}\}, l=1, 2.
$$
In the data generation, ``$\cdot$" is a generic notation, which can be ``$t$" for the treatment group and ``$c$" for the control group. In our simulation, $F_{l(t)}(\cdot)$ is the CDF of an exponential distribution with shape parameters $\lambda_l^{(t)},$ which is 0.015 for $l=1$ and 0.02 for $l=2.$

We consider three simulation settings. In the first setting, we simulate the null case by letting $F_{l(c)}(\cdot)$ to be the same as $F_{l(t)}(\cdot)$ in the treatment group.
The second setting is the same as the first, except that $(\lambda_1^{(c)},\lambda_2^{(c)}) = (0.021,0.029)$, representing a proportional hazards alternative. In the third setting, $F_{l(c)}(\cdot), l=1, 2, 3$ are the CDFs of piece-wise exponential distributions:
\begin{eqnarray*}
\lambda_1^{(c)}(s) &=& 0.015 + 0.006I(s\ge 5);\\
\lambda_2^{(c)}(s) &=& 0.020 + 0.009I(s\ge 5)
\end{eqnarray*}
to investigate the performance of proposed inference procedure with a delayed treatment effect. 
$F_{C_l(\cdot)}(\cdot)$ is set as the CDF of an exponential distribution with shape parameters $\lambda_{C_l}^{(\cdot)},$ which is 0.015 for $l=1$ and 0.02 for $l=2.$ We examine varying levels of correlation between the censoring times for the two endpoints by setting $\rho$ at 0.25, 0.5 and 0.75.
Depending on the simulation setting, the censoring rate of the truncated event times for these endpoints typically ranges from 20\% to 50\%. Three sets of sample sizes considered include $(N_t,N_c)=(100,100)$, $(200,200)$, and $(400,400)$. We focus on estimating the win probabilities of treatment and control, and the win ratio with $\tau = 36$ and $\zeta_1=\zeta_2=\zeta_3=\zeta$, where $\zeta=0$ or 6.

The results are summarized in Table \ref{tab:CS_EXP1}. All three methods show similarly small biases, indicating that the validity of the estimator is not substantially affected by the use of induced common censoring. In terms of variance, the first two methods yield comparable results, suggesting that estimating the survival function of the induced common censoring time does not materially impact the accuracy of win probability and win ratio estimation. In contrast, the third method, which uses IPCW adjustment based on the true joint distribution of censoring times, generally produces lower variance. This highlights a modest efficiency loss associated with inducing common censoring. The efficiency loss is more pronounced when the censoring times of the two endpoints are weakly correlated (e.g., $\rho=$ 0 or 0.25), and less so when the correlation is stronger (e.g., $\rho=0.75$). This is expected, as a higher correlation implies a smaller difference between $C_1$ (or $C_2$) and $C_1\wedge C_2.$ Moreover, this efficiency gap diminishes as the sample size increases. Overall, the results support that the proposed estimator with induced common censoring remains valid and incurs only modest efficiency loss.
}

\begin{table}[H]
  \centering  \footnotesize
  \caption{ Simulation results to evaluate the impact of inducing common censoring based on multivariate censoring times.}
    \begin{tabular}{c|c|c|c|c|c|c|c|c|c|c|c|c}
    \hline
    \multirow{2}{*}{Case} & \multirow{2}{*}{$\rho$} & \multirow{2}{*}{$n$} & \multirow{2}{*}{$\zeta$} & \multicolumn{3}{c|}{WR} & \multicolumn{3}{c|}{WR (TC)} & \multicolumn{3}{c}{WR (TJ)} \\
\cline{5-13}          &       &       &       & TRUE  & \multicolumn{1}{c|}{BIAS} & \multicolumn{1}{c|}{Var} & TRUE  & \multicolumn{1}{c|}{BIAS} & \multicolumn{1}{c|}{Var} & TRUE  & \multicolumn{1}{c|}{BIAS} & \multicolumn{1}{c}{Var} \\
    \hline
    \multirow{18}{*}{I} & \multirow{6}{*}{0.25} & \multirow{2}{*}{(100,100)} & 0     & 1.000 & 0.016 & 0.068 & 1.000 & 0.014 & 0.066 & 1.000 & 0.012 & 0.050 \\
\cline{4-13}          &       &       & 6     & 1.000 & 0.019 & 0.077 & 1.000 & 0.017 & 0.075 & 1.000 & 0.016 & 0.058 \\
\cline{3-13}          &       & \multirow{2}{*}{(200,200)} & 0     & 1.000 & 0.016 & 0.033 & 1.000 & 0.016 & 0.033 & 1.000 & 0.011 & 0.025 \\
\cline{4-13}          &       &       & 6     & 1.000 & 0.016 & 0.037 & 1.000 & 0.017 & 0.038 & 1.000 & 0.011 & 0.028 \\
\cline{3-13}          &       & \multirow{2}{*}{(400,400)} & 0     & 1.000 & 0.008 & 0.016 & 1.000 & 0.008 & 0.016 & 1.000 & 0.006 & 0.012 \\
\cline{4-13}          &       &       & 6     & 1.000 & 0.010 & 0.018 & 1.000 & 0.010 & 0.018 & 1.000 & 0.007 & 0.014 \\
\cline{2-13}          & \multirow{6}{*}{0.5} & \multirow{2}{*}{(100,100)} & 0     & 1.000 & 0.014 & 0.063 & 1.000 & 0.012 & 0.062 & 1.000 & 0.010 & 0.048 \\
\cline{4-13}          &       &       & 6     & 1.000 & 0.018 & 0.071 & 1.000 & 0.017 & 0.071 & 1.000 & 0.014 & 0.056 \\
\cline{3-13}          &       & \multirow{2}{*}{(200,200)} & 0     & 1.000 & 0.014 & 0.030 & 1.000 & 0.014 & 0.030 & 1.000 & 0.010 & 0.024 \\
\cline{4-13}          &       &       & 6     & 1.000 & 0.015 & 0.034 & 1.000 & 0.016 & 0.035 & 1.000 & 0.010 & 0.027 \\
\cline{3-13}          &       & \multirow{2}{*}{(400,400)} & 0     & 1.000 & 0.007 & 0.015 & 1.000 & 0.007 & 0.015 & 1.000 & 0.006 & 0.012 \\
\cline{4-13}          &       &       & 6     & 1.000 & 0.009 & 0.017 & 1.000 & 0.009 & 0.017 & 1.000 & 0.007 & 0.014 \\
\cline{2-13}          & \multirow{6}{*}{0.75} & \multirow{2}{*}{(100,100)} & 0     & 1.000 & 0.013 & 0.058 & 1.000 & 0.011 & 0.058 & 1.000 & 0.012 & 0.047 \\
\cline{4-13}          &       &       & 6     & 1.000 & 0.015 & 0.067 & 1.000 & 0.014 & 0.067 & 1.000 & 0.013 & 0.055 \\
\cline{3-13}          &       & \multirow{2}{*}{(200,200)} & 0     & 1.000 & 0.015 & 0.029 & 1.000 & 0.015 & 0.029 & 1.000 & 0.010 & 0.023 \\
\cline{4-13}          &       &       & 6     & 1.000 & 0.015 & 0.033 & 1.000 & 0.015 & 0.033 & 1.000 & 0.009 & 0.027 \\
\cline{3-13}          &       & \multirow{2}{*}{(400,400)} & 0     & 1.000 & 0.006 & 0.014 & 1.000 & 0.006 & 0.014 & 1.000 & 0.007 & 0.012 \\
\cline{4-13}          &       &       & 6     & 1.000 & 0.008 & 0.016 & 1.000 & 0.008 & 0.016 & 1.000 & 0.008 & 0.013 \\
    \hline
    \multirow{18}{*}{II} & \multirow{6}{*}{0.25} & \multirow{2}{*}{(100,100)} & 0     & 1.402 & 0.020 & 0.121 & 1.402 & 0.019 & 0.120 & 1.402 & 0.018 & 0.088 \\
\cline{4-13}          &       &       & 6     & 1.417 & 0.055 & 0.143 & 1.417 & 0.053 & 0.141 & 1.417 & 0.053 & 0.109 \\
\cline{3-13}          &       & \multirow{2}{*}{(200,200)} & 0     & 1.402 & 0.022 & 0.059 & 1.402 & 0.022 & 0.059 & 1.402 & 0.016 & 0.046 \\
\cline{4-13}          &       &       & 6     & 1.417 & 0.052 & 0.070 & 1.417 & 0.053 & 0.071 & 1.417 & 0.046 & 0.054 \\
\cline{3-13}          &       & \multirow{2}{*}{(400,400)} & 0     & 1.402 & 0.013 & 0.029 & 1.402 & 0.013 & 0.029 & 1.402 & 0.008 & 0.022 \\
\cline{4-13}          &       &       & 6     & 1.417 & 0.045 & 0.035 & 1.417 & 0.046 & 0.035 & 1.417 & 0.039 & 0.027 \\
\cline{2-13}          & \multirow{6}{*}{0.5} & \multirow{2}{*}{(100,100)} & 0     & 1.402 & 0.019 & 0.110 & 1.402 & 0.018 & 0.110 & 1.402 & 0.017 & 0.085 \\
\cline{4-13}          &       &       & 6     & 1.417 & 0.052 & 0.129 & 1.417 & 0.053 & 0.129 & 1.417 & 0.051 & 0.105 \\
\cline{3-13}          &       & \multirow{2}{*}{(200,200)} & 0     & 1.402 & 0.021 & 0.054 & 1.402 & 0.021 & 0.054 & 1.402 & 0.015 & 0.044 \\
\cline{4-13}          &       &       & 6     & 1.417 & 0.051 & 0.065 & 1.417 & 0.053 & 0.066 & 1.417 & 0.045 & 0.053 \\
\cline{3-13}          &       & \multirow{2}{*}{(400,400)} & 0     & 1.402 & 0.009 & 0.028 & 1.402 & 0.009 & 0.028 & 1.402 & 0.007 & 0.023 \\
\cline{4-13}          &       &       & 6     & 1.417 & 0.041 & 0.033 & 1.417 & 0.041 & 0.034 & 1.417 & 0.038 & 0.027 \\
\cline{2-13}          & \multirow{6}{*}{0.75} & \multirow{2}{*}{(100,100)} & 0     & 1.402 & 0.020 & 0.105 & 1.402 & 0.018 & 0.105 & 1.402 & 0.017 & 0.085 \\
\cline{4-13}          &       &       & 6     & 1.417 & 0.052 & 0.125 & 1.417 & 0.050 & 0.124 & 1.417 & 0.050 & 0.104 \\
\cline{3-13}          &       & \multirow{2}{*}{(200,200)} & 0     & 1.402 & 0.020 & 0.052 & 1.402 & 0.019 & 0.052 & 1.402 & 0.014 & 0.044 \\
\cline{4-13}          &       &       & 6     & 1.417 & 0.049 & 0.063 & 1.417 & 0.050 & 0.064 & 1.417 & 0.045 & 0.052 \\
\cline{3-13}          &       & \multirow{2}{*}{(400,400)} & 0     & 1.402 & 0.010 & 0.025 & 1.402 & 0.010 & 0.026 & 1.402 & 0.009 & 0.022 \\
\cline{4-13}          &       &       & 6     & 1.417 & 0.044 & 0.031 & 1.417 & 0.044 & 0.032 & 1.417 & 0.041 & 0.027 \\
    \hline
    \multirow{18}{*}{III} & \multirow{6}{*}{0.25} & \multirow{2}{*}{(100,100)} & 0     & 1.315 & 0.015 & 0.107 & 1.315 & 0.013 & 0.107 & 1.315 & 0.012 & 0.078 \\
\cline{4-13}          &       &       & 6     & 1.321 & 0.043 & 0.127 & 1.321 & 0.040 & 0.125 & 1.321 & 0.039 & 0.096 \\
\cline{3-13}          &       & \multirow{2}{*}{(200,200)} & 0     & 1.315 & 0.018 & 0.052 & 1.315 & 0.017 & 0.053 & 1.315 & 0.013 & 0.041 \\
\cline{4-13}          &       &       & 6     & 1.321 & 0.039 & 0.061 & 1.321 & 0.039 & 0.062 & 1.321 & 0.033 & 0.047 \\
\cline{3-13}          &       & \multirow{2}{*}{(400,400)} & 0     & 1.315 & 0.010 & 0.026 & 1.315 & 0.010 & 0.026 & 1.315 & 0.007 & 0.020 \\
\cline{4-13}          &       &       & 6     & 1.321 & 0.034 & 0.031 & 1.321 & 0.035 & 0.032 & 1.321 & 0.029 & 0.024 \\
\cline{2-13}          & \multirow{6}{*}{0.5} & \multirow{2}{*}{(100,100)} & 0     & 1.315 & 0.015 & 0.097 & 1.315 & 0.014 & 0.097 & 1.315 & 0.011 & 0.075 \\
\cline{4-13}          &       &       & 6     & 1.321 & 0.042 & 0.114 & 1.321 & 0.042 & 0.114 & 1.321 & 0.037 & 0.091 \\
\cline{3-13}          &       & \multirow{2}{*}{(200,200)} & 0     & 1.315 & 0.017 & 0.048 & 1.315 & 0.016 & 0.048 & 1.315 & 0.012 & 0.039 \\
\cline{4-13}          &       &       & 6     & 1.321 & 0.037 & 0.056 & 1.321 & 0.039 & 0.057 & 1.321 & 0.032 & 0.046 \\
\cline{3-13}          &       & \multirow{2}{*}{(400,400)} & 0     & 1.315 & 0.008 & 0.024 & 1.315 & 0.008 & 0.024 & 1.315 & 0.007 & 0.019 \\
\cline{4-13}          &       &       & 6     & 1.321 & 0.032 & 0.029 & 1.321 & 0.032 & 0.029 & 1.321 & 0.030 & 0.023 \\
\cline{2-13}          & \multirow{6}{*}{0.75} & \multirow{2}{*}{(100,100)} & 0     & 1.315 & 0.014 & 0.092 & 1.315 & 0.012 & 0.092 & 1.315 & 0.011 & 0.075 \\
\cline{4-13}          &       &       & 6     & 1.321 & 0.040 & 0.109 & 1.321 & 0.037 & 0.108 & 1.321 & 0.037 & 0.091 \\
\cline{3-13}          &       & \multirow{2}{*}{(200,200)} & 0     & 1.315 & 0.016 & 0.046 & 1.315 & 0.016 & 0.046 & 1.315 & 0.011 & 0.038 \\
\cline{4-13}          &       &       & 6     & 1.321 & 0.037 & 0.055 & 1.321 & 0.037 & 0.055 & 1.321 & 0.032 & 0.045 \\
\cline{3-13}          &       & \multirow{2}{*}{(400,400)} & 0     & 1.315 & 0.007 & 0.023 & 1.315 & 0.007 & 0.023 & 1.315 & 0.006 & 0.019 \\
\cline{4-13}          &       &       & 6     & 1.321 & 0.032 & 0.028 & 1.321 & 0.032 & 0.028 & 1.321 & 0.029 & 0.023 \\
    \hline
    \end{tabular}
  \label{tab:CS_EXP1}
\end{table}

{\section{Additional Figures and Tables}

\begin{table}[H]
  \centering
  \caption{Simulation results on bias (BIAS) in estimating the win probability for treatment and control ($\pi_t^{(\text{no})}$ and $\pi_c^{(\text{no})}$, respectively), and bias (BIAS) in estimating win ratio, average analytical standard error estimate of $\log(\widehat{WR}^{(\text{no})})$ (ASE), empirical standard error of $\log(\widehat{WR}^{(\text{no})})$ (ESE) and empirical coverage probability (CP) of 95\% CIs for the naive win ratio estimator without IPCW adjustment across 5000 replicates with $\zeta = 0$ and $\tau\in \{18, 36\}$ in settings of exponential distributions.}
\begin{tabular}{c|c|c|c|r|c|r|c|c|c|c|c}
\hline
& & & \multicolumn{2}{c|}{$\pi_t^{(\text{no})}$} & \multicolumn{2}{c|}{$\pi_c^{(\text{no})}$} & \multicolumn{5}{c}{WR$^{(\text{no})}$} \\
\cline{4-12}  Setting  & $\tau$ & $(N_t,N_c)$ & TRUE   & BIAS  & TRUE   & BIAS  & TRUE   & BIAS  & ASE   & ESE   & CP \\
\hline
I & 18 & (100,100) & 0.447 & -0.084 & 0.447 & -0.084 & 1.000 & 0.017 & 0.192 & 0.195 & 0.944 \\ 
&       & (200,200) & 0.447 & -0.083 & 0.447 & -0.085 & 1.000 & 0.012 & 0.136 & 0.134 & 0.946 \\ 
&       & (400,400) & 0.447 & -0.084 & 0.447 & -0.084 & 1.000 & 0.004 & 0.096 & 0.096 & 0.947 \\ 
& 36 & (100,100) & 0.493 & -0.112 & 0.493 & -0.111 & 1.000 & 0.014 & 0.180 & 0.182 & 0.948 \\
&       & (200,200) & 0.493 & -0.111 & 0.493 & -0.112 & 1.000 & 0.011 & 0.128 & 0.126 & 0.949 \\
&       & (400,400) & 0.493 & -0.112 & 0.493 & -0.112 & 1.000 & 0.003 & 0.090 & 0.090 & 0.948 \\ 
\hline
II & 18 & (100,100) & 0.519 & -0.095 & 0.397 & -0.068 & 1.307 & 0.009 & 0.188 & 0.190 & 0.950 \\
&       & (200,200) & 0.519 & -0.094 & 0.397 & -0.069 & 1.307 & 0.003 & 0.133 & 0.131 & 0.952 \\
&       & (400,400) & 0.519 & -0.095 & 0.397 & -0.069 & 1.307 & -0.011 & 0.094 & 0.094 & 0.948 \\
& 36 & (100,100) &  0.571 & -0.127 & 0.420 & -0.082 & 1.360 & -0.022 & 0.179 & 0.179 & 0.948 \\ 
&       & (200,200) & 0.571 & -0.126 & 0.420 & -0.083 & 1.360 & -0.027 & 0.127 & 0.125 & 0.946 \\
&       & (400,400) & 0.571 & -0.127 & 0.420 & -0.082 & 1.360 & -0.038 & 0.090 & 0.089 & 0.938 \\ 
\hline
III & 18 & (100,100) & 0.497 & -0.100 & 0.414 & -0.071 & 1.198 & -0.019 & 0.189 & 0.191 & 0.944 \\
&       & (200,200) & 0.497 & -0.099 & 0.414 & -0.072 & 1.198 & -0.025 & 0.134 & 0.132 & 0.947 \\ 
&       & (400,400) & 0.497 & -0.100 & 0.414 & -0.071 & 1.198 & -0.036 & 0.095 & 0.094 & 0.935 \\  
& 36 & (100,100) & 0.556 & -0.136 & 0.435 & -0.083 & 1.278 & -0.063 & 0.179 & 0.180 & 0.934 \\
&       & (200,200) & 0.556 & -0.135 & 0.435 & -0.084 & 1.278 & -0.067 & 0.127 & 0.126 & 0.925 \\ 
&       & (400,400) & 0.556 & -0.136 & 0.435 & -0.084 & 1.278 & -0.078 & 0.090 & 0.089 & 0.886 \\
\hline
\end{tabular}
\label{tab:SEXP_1}
\end{table}

\begin{table}[H]
  \centering
  \caption{Simulation results on bias (BIAS) in estimating the win probability for treatment and control ($\pi_t^{(\text{no})}$ and $\pi_c^{(\text{no})}$, respectively), and bias (BIAS) in estimating win ratio, average analytical standard error estimate of $\log(\widehat{WR}^{(\text{no})})$ (ASE), empirical standard error of $\log(\widehat{WR}^{(\text{no})})$ (ESE) and empirical coverage probability (CP) of 95\% CIs for the naive win ratio estimator without IPCW adjustment across 5000 replicates with different combinations of $\tau$s and $\zeta$s in different settings of exponential distributions ($N_t=N_c=200$).}
    \begin{tabular}{c|c|r|c|r|c|r|c|c|c|c|c}
    \hline
    \multirow{2}{*}{} & &  & \multicolumn{2}{c|}{$\pi_t^{(\text{no})}$} & \multicolumn{2}{c|}{$\pi_c^{(\text{no})}$} & \multicolumn{5}{c}{WR$^{(\text{no})}$} \\
\cline{4-12}      Setting    & $\tau$ &   $\zeta$    & TRUE   & BIAS  & TRUE   & BIAS  & TRUE   & BIAS  & ASE   & ESE   & CP \\
    \hline
    I & 18 & 0     & 0.447 & -0.083 & 0.447 & -0.085 & 1.000 & 0.012 & 0.136 & 0.134 & 0.946 \\
&       & 2     & 0.424 & -0.089 & 0.424 & -0.091 & 1.000 & 0.014 & 0.142 & 0.142 & 0.946 \\ 
&       & 4     & 0.393 & -0.091 & 0.393 & -0.092 & 1.000 & 0.016 & 0.150 & 0.149 & 0.945 \\
&       & 6     & 0.356 & -0.088 & 0.356 & -0.090 & 1.000 & 0.018 & 0.160 & 0.160 & 0.946 \\
& 36 & 0     & 0.493 & -0.111 & 0.493 & -0.112 & 1.000 & 0.011 & 0.128 & 0.126 & 0.949 \\
&       & 2     & 0.487 & -0.127 & 0.487 & -0.128 & 1.000 & 0.012 & 0.130 & 0.129 & 0.948 \\
&       & 4     & 0.478 & -0.140 & 0.478 & -0.142 & 1.000 & 0.013 & 0.134 & 0.133 & 0.946 \\
&       & 6     & 0.466 & -0.151 & 0.466 & -0.152 & 1.000 & 0.013 & 0.138 & 0.137 & 0.947 \\ 
\hline
II & 18 & 0     & 0.519 & -0.094 & 0.397 & -0.069 & 1.307 & 0.003 & 0.133 & 0.131 & 0.952 \\
&       & 2     & 0.496 & -0.103 & 0.378 & -0.076 & 1.313 & 0.003 & 0.139 & 0.138 & 0.952 \\ 
&       & 4     & 0.465 & -0.107 & 0.353 & -0.080 & 1.316 & 0.010 & 0.147 & 0.146 & 0.955 \\ 
&       & 6     & 0.424 & -0.105 & 0.321 & -0.079 & 1.320 & 0.017 & 0.156 & 0.156 & 0.951 \\
& 36 & 0     & 0.571 & -0.126 & 0.420 & -0.083 & 1.360 & -0.027 & 0.127 & 0.125 & 0.946 \\
&       & 2     & 0.567 & -0.145 & 0.415 & -0.098 & 1.367 & -0.024 & 0.129 & 0.128 & 0.945 \\
&       & 4     & 0.559 & -0.161 & 0.408 & -0.111 & 1.371 & -0.018 & 0.133 & 0.131 & 0.947 \\
&       & 6     & 0.547 & -0.174 & 0.398 & -0.121 & 1.376 & -0.012 & 0.137 & 0.136 & 0.951 \\
\hline
III & 18 & 0     & 0.497 & -0.099 & 0.414 & -0.072 & 1.198 & -0.025 & 0.134 & 0.132 & 0.947 \\
&       & 2     & 0.471 & -0.106 & 0.395 & -0.080 & 1.193 & -0.022 & 0.140 & 0.139 & 0.947 \\
&       & 4     & 0.437 & -0.106 & 0.369 & -0.084 & 1.183 & -0.012 & 0.148 & 0.148 & 0.950 \\
&       & 6     & 0.394 & -0.102 & 0.336 & -0.083 & 1.172 & -0.001 & 0.158 & 0.159 & 0.950 \\
& 36 & 0     & 0.556 & -0.135 & 0.435 & -0.084 & 1.278 & -0.067 & 0.127 & 0.126 & 0.925 \\ 
&       & 2     & 0.550 & -0.153 & 0.430 & -0.100 & 1.280 & -0.063 & 0.130 & 0.128 & 0.928 \\
&       & 4     & 0.542 & -0.167 & 0.423 & -0.114 & 1.281 & -0.057 & 0.133 & 0.132 & 0.931 \\
&       & 6     & 0.530 & -0.179 & 0.413 & -0.125 & 1.284 & -0.050 & 0.137 & 0.136 & 0.937 \\  
    \hline
    \end{tabular}
  \label{tab:SEXP_2}
\end{table}

\newpage
\begin{figure}[H]
    \centering 
    \includegraphics[width=\textwidth]{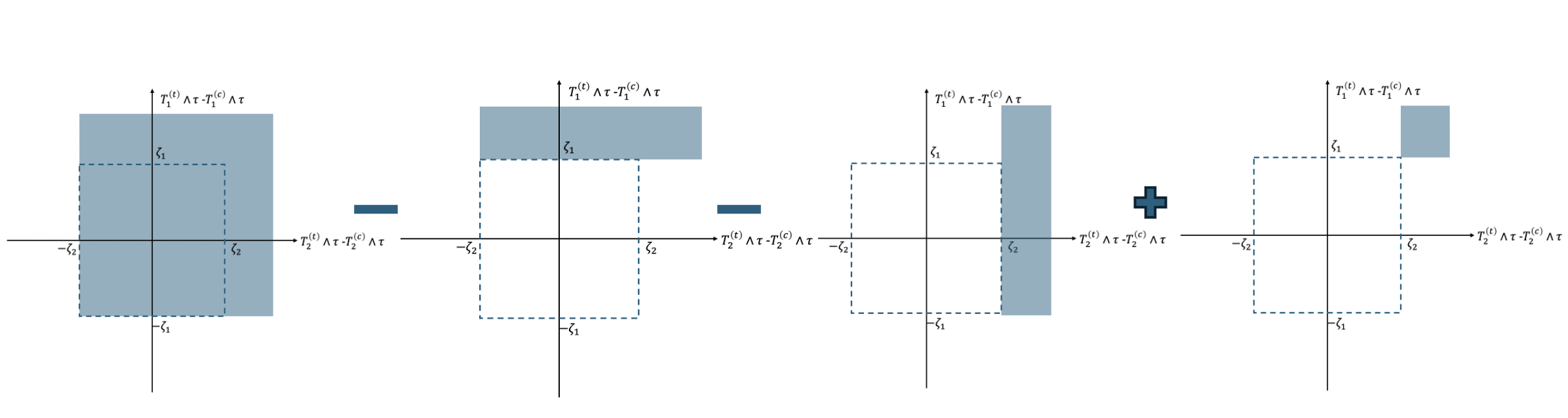}
    \caption{ Visualization of the inclusion-exclusion decomposition used in estimating $\pi_{t3}$.}
    \label{fig:S1}
\end{figure}
}

\newpage
\begin{figure}
    \centering
    \includegraphics[width = \textwidth]{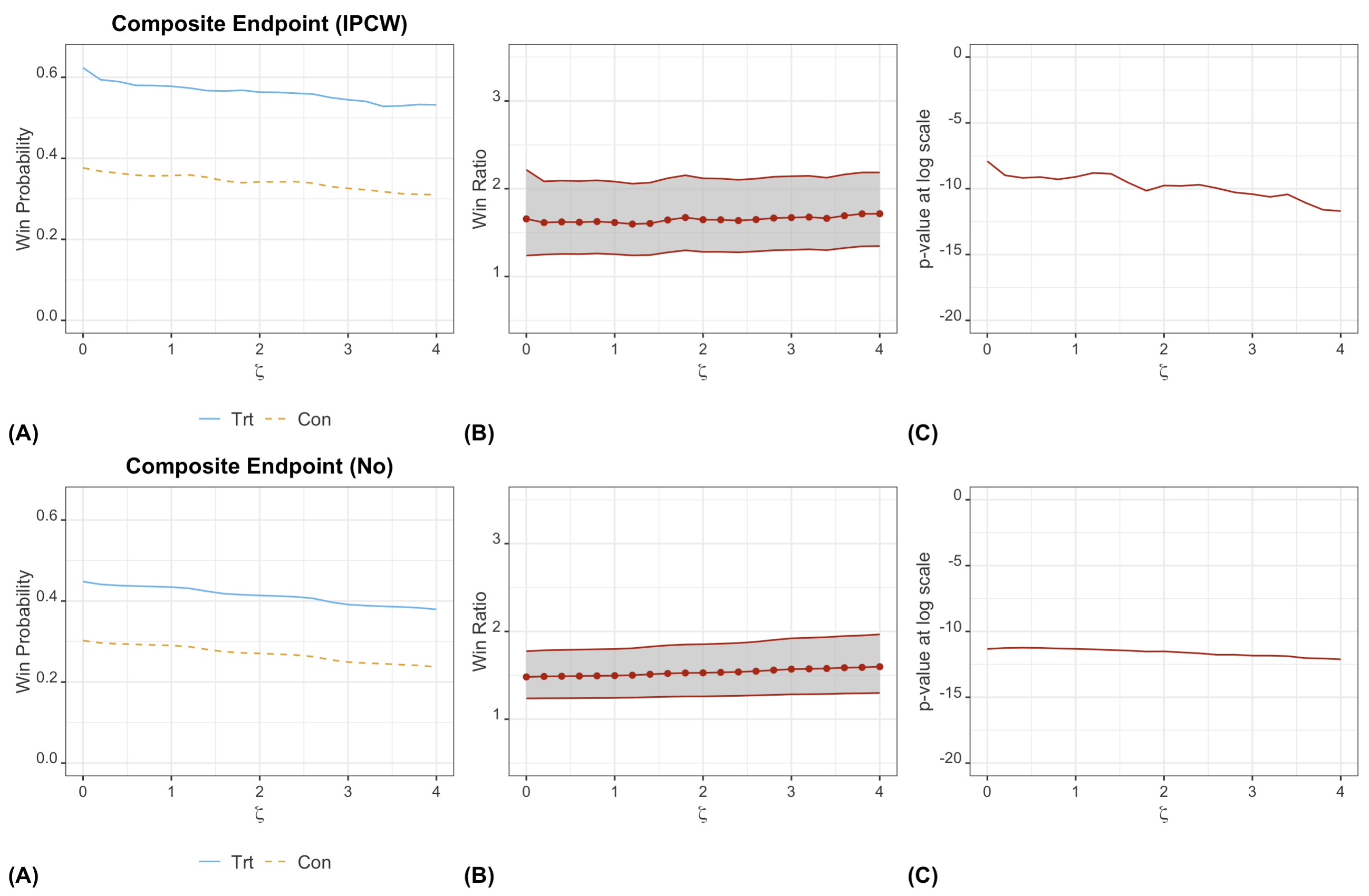}
    \caption{{ Comparison of analysis results with and without IPCW adjustment in JAVELIN Renal 101 trial. Panel A: win probabilities of treatment and control; Panel B: the estimated win ratios with 95\% CI; Panel C: $p$ value for testing the treatment effect based on the win ratio (log scale).}}
    \label{fig:S2}
\end{figure}

\bibliographystyle{biom}
\bibliography{WINS}

%% file: SIM_revision/term_explain.tex
The terms in the estimator $\widehat{\pi}_{tl}$ arise from the application of an inclusion-exclusion principle to account for the ties with the outcomes $\{T_k\}_{k=1}^{l-1}$. Figure S1 of the supplementary material provides a visual illustration of this principle for $l = 3$. Specifically, the region of interest for the ties, defined by $\{-\zeta_2 \le T_2^{(t)}\wedge \tau - T_2^{(c)}\wedge \tau \le \zeta_2,-\zeta_1 \le T_1^{(t)}\wedge \tau - T_1^{(c)}\wedge \tau \le \zeta_1\}$, is decomposed into the inclusion and exclusion of four regions. The first panel corresponds to the inclusive region, $\{T_2^{(t)}\wedge \tau - T_2^{(c)}\wedge \tau \ge -\zeta_2, T_1^{(t)}\wedge \tau - T_1^{(c)}\wedge \tau \ge -\zeta_1\}$. To isolate the target region, we subtract the areas $\{T_1^{(t)}\wedge \tau - T_1^{(c)}\wedge \tau \ge \zeta_1\}$ and $\{T_2^{(t)}\wedge \tau - T_2^{(c)}\wedge \tau \ge \zeta_2\}$ shown in the second and third panels, respectively. The doubly excluded region, $\{T_2^{(t)}\wedge \tau - T_2^{(c)}\wedge \tau \ge \zeta_2, T_1^{(t)}\wedge \tau - T_1^{(c)}\wedge \tau \ge \zeta_1\},$ is added back in the final panel. This visual breakdown aligns with the structure of the estimator, where each term corresponds to one of the shaded regions.

%% file: SIM_revision/sensitivity_analyses_zeta_response.tex
Additionally, the results show that when $\tau$ is not too small, the statistical power can be quite robust to the choice of $\zeta$. For example, in Table 3, with a moderate sample size of 200 per group, the empirical rejection rate decreases from 0.35 to 0.25 as $\zeta$ increases from 0 to 6, when $\tau = 18$. In contrast, when $\tau = 36$, the empirical rejection rate remains consistent around 0.54 regardless of the value of $\zeta$. 

%% file: SIM_revision/discussion_WR_without_IPCW.tex
We also compared the proposed method with the naive win ratio estimator without IPCW adjustment. Results are presented in Table 3 of the manuscript and Tables S7 and S8 of the supplementary materials. Without IPCW adjustment, the naïve estimator exhibits bias in estimating the win probability for either treatment or control, which in turn induces bias in estimating the win statistic—defined as the contrast between these two win probabilities, particularly when a nonzero treatment effect is present. In terms of inference, the coverage probability of confidence intervals for the win statistic deteriorates with increasing sample size due to this bias. For example, as shown in Table S6, the coverage probability for Setting III decreases from 0.934 to 0.886 as the sample size increases from 100 to 400 per group. While these methods perform similarly in hypothesis testing in general, the proposed IPCW-adjusted approach demonstrates greater power for detecting a delayed treatment effect. These simulation results underscore the advantages of our proposed method and highlight the importance of incorporating IPCW adjustment.

%% file: SIM_revision/discussion_Weibull.tex
To further evaluate the robustness of our proposed method under alternative survival distributions, we conducted additional simulation studies using data generated from Weibull distributions. Details are provided in Section S5.1 of the supplementary materials. The results, summarized in Tables S1–S5, demonstrate that the proposed method performs consistently well. Similar to findings based on exponential distributions, the proposed approach exhibits minimal bias, accurate standard error estimation, and correct coverage probabilities for the constructed confidence intervals. Compared to the log-rank test, which only accounts for time to the first event, the proposed tests offer substantially greater power.

%% file: SIM_revision/discussion_common_censoring.tex
Finally, we conducted a simulation study to evaluate the performance of the proposed method with an induced common censoring, when the censoring time varies with different survival outcome. Details of the simulation design are provided in Section S5.2 of the supplementary materials, with results summarized in Table S6. The proposed estimator with induced common censoring is nearly unbiased, and the associated efficiency loss is modest when compared to the IPCW-adjusted estimator based on the true joint distribution of multivariate censoring times. 

%% file: SIM_revision/real_compare_HR_RMST.tex
Furthermore, in the special case where $\zeta=0$, we compared the result of the proposed method with two established approaches: the Cox proportional hazards model and the restricted mean survival time (RMST)–based comparison for time to the first event. The estimated hazard ratio from Cox regression was 1.54 (95\% CI: [1.31, 1.80]) with a p-value $<$ 0.0001, while the RMST ratio between the treatment and control groups was 1.31 (95\% CI: [1.18, 1.45]) with a p-value $<$ 0.0001. These findings are consistent with those from the proposed method at $\zeta=0,$ where the estimated win ratio was 1.66 (95\% CI: [1.24, 2.22]) with a significant p-value of 0.0004. Collectively, these results suggest a beneficial treatment effect on the composite endpoint. In the context of decision making in clinical practice, the proposed method provided not only a statistically significant evidence for the presence of treatment benefit, but also a transparent interpretation of the estimated win ratio integrating the treatment effects on both OS and PFS.

%% file: SIM_revision/real_WR_without_IPCW.tex
We also obtained the naive estimator of the win ratio without IPCW adjustment, with results shown in Figure S2 of the supplementary materials. The estimated win probabilities without IPCW adjustment were consistently lower than their IPCW-adjusted counterparts. Interestingly, the test for treatment effect appeared more significant using the naive estimator—may due to the fact that there is no need to account for the variability introduced by right-censoring. However, it is important to note that a more significant p-value does not necessarily equal to a ``better'' result. Interpreting the magnitude of treatment benefit relies on the estimated win probabilities, which can be substantially biased without appropriate IPCW adjustment as our simulation study demonstrates.

%% file: SIM_revision/discussionnew1.tex
When $\tau$ is relatively small, the choice of $\zeta$ can have a greater impact on the analysis result; specifically, a larger margin $\zeta$ may reduce statistical power. However, when $\tau$ is sufficiently large, the statistical power is quite robust to the value of $\zeta$.

%% file: SIM_revision/discussion_choosing_zeta.tex
Notably, the equivalence margin $\zeta$ should be chosen to reflect the minimum clinically important difference, i.e., a difference below which the treatment effect would be considered clinically negligible. The selection of $\zeta$ should be primarily guided by practical clinical considerations and informed by established benchmarks. For example, as noted by Panageas et al. \citep{panageas2007you}, the date of disease progression is typically determined through radiological evaluations at scheduled intervals and thus only approximates the true progression time. Because progression may occur between evaluations, defining $\zeta$ based on the typical time between assessments can help avoid overestimating treatment efficacy when analyzing progression-free survival. For overall survival, a relative improvement of 20\% in median survival is generally considered clinically meaningful \citep{ellis2014american}. Therefore, a reasonable choice for $\zeta$ in this context is 20\% of the median survival time for the indication of interest. Other factors may also influence the selection of $\zeta.$ For instance, if a therapy offers lower toxicity or improved tolerability compared to standard treatments, a smaller efficacy improvement may still be considered clinically relevant and acceptable.